\documentclass[12pt]{article}
\usepackage{amsmath,amssymb,amsthm,amsxtra,overpic,bbm,bm,epsfig,sub
figure}
\usepackage{mathrsfs}
\usepackage{graphicx}
\usepackage{color}
\usepackage{comment}
\usepackage{epstopdf}
\usepackage{float}
\usepackage{slashed,stmaryrd}

\def\thefootnote{\fnsymbol{footnote}}

\begin{document}

\vspace{0.2cm}

\begin{center}
{\Large\bf Sum rules and asymptotic behaviors of neutrino mixing \\
in dense matter}
\end{center}

\vspace{0.2cm}

\begin{center}
{\bf Zhi-zhong Xing$^{1, 2, 3}$} \footnote{E-mail:
xingzz@ihep.ac.cn}
and {\bf Jing-yu Zhu$^{1, 2}$} \footnote{E-mail:
zhujingyu@ihep.ac.cn}
\\
{\small $^{1}$Institute of High Energy Physics, Chinese Academy of
Sciences, Beijing 100049, China \\
$^{2}$School of Physical Sciences, University of Chinese Academy of
Sciences,
Beijing 100049, China \\
$^{3}$Center for High Energy Physics, Peking University, Beijing
100871, China}
\end{center}

\vspace{1.5cm}
\begin{abstract}
It has proved convenient to define the {\it effective} lepton flavor mixing matrix
$\widetilde{U}$ and neutrino mass-squared differences $\widetilde{\Delta}^{}_{ji}
\equiv \widetilde{m}^2_j - \widetilde{m}^2_i$ (for $i,j =1,2,3$) to describe the
phenomena of neutrino mixing and flavor oscillations in a medium, but the prerequisite
is to establish direct and transparent relations between these effective
quantities and their fundamental counterparts in vacuum.
With the help of two sets of sum rules for $\widetilde{U}$ and
$\widetilde{\Delta}^{}_{ji}$, we derive new and exact formulas for moduli of the nine
elements of $\widetilde{U}$ and the sides of its three Dirac unitarity triangles
in the complex plane. The asymptotic behaviors of $|\widetilde{U}^{}_{\alpha i}|^2$
and $\widetilde{\Delta}^{}_{ji}$ (for $\alpha = e, \mu, \tau$ and $i,j =1,2,3$) in
very dense matter (namely, allowing the matter parameter
$A = 2\sqrt{2} ~ G^{}_{\rm F} N^{}_e E$ to mathematically approach infinity) are
{\it analytically} unraveled for the first time, and in this connection the confusion
associated with the parameter redundancy of $\widetilde{\theta}^{}_{12}$, $\widetilde{\theta}^{}_{13}$, $\widetilde{\theta}^{}_{23}$
and $\widetilde{\delta}$ in the standard parametrization of $\widetilde{U}$ is
clarified.
\end{abstract}
\begin{flushleft}
\hspace{0.8cm} PACS number(s): 14.60.Pq, 25.30.Pt
\end{flushleft}

\def\thefootnote{\arabic{footnote}}
\setcounter{footnote}{0}

\newpage

\section{Introduction}

When a neutrino beam travels in a medium, its electron-flavor component
undergoes some forward coherent scattering with the electrons in this
medium via the weak charged-current interactions, leading to a nontrivial
modification of the behaviors of neutrino
oscillations \cite{Wolfenstein:1977ue, Mikheev:1986gs, Mikheev:1986wj}.
Such matter effects have played very important roles in solving the long-standing
solar neutrino problem and in explaining current atmospheric and long-baseline
accelerator neutrino oscillation data \cite{Tanabashi:2018oca}, and they are even
expected to have an appreciable impact on the sensitivity of a medium-baseline
JUNO-like reactor antineutrino oscillation experiment \cite{Li:2016txk,Li:2018jgd}.
A lot of efforts have been made in the past decades to formulate matter
effects on neutrino oscillations, and recently some interest has been shown in
going beyond Freund's analytical approximations \cite{Freund:2001pn}
to reformulate probabilities of neutrino oscillations with weak or strong
terrestrial matter contamination (see, e.g., Refs.
\cite{Akhmedov:2004ny,Blennow:2013rca,Xu:2015kma,Xing:2016ymg,Denton:2016wmg,Li:2016pzm,Ioannisian:2018qwl,
Huang:2018ufu,Denton:2018fex,Petcov:2018zka}),
or in describing matter effects on neutrino
mixing and CP violation with the help of a language similar to the
renormalization-group equations (see, e.g., Refs. \cite{Zhou:2016luk,Chiu:2017ckv,Xing:2018lob,Wang:2019yfp}).

With the help of two sets of sum rules for the {\it effective} neutrino mass-squared
differences $\widetilde{\Delta}^{}_{ji} \equiv \widetilde{m}^2_j - \widetilde{m}^2_i$
(for $i,j =1,2,3$) and the {\it effective} Pontecorvo-Maki-Nakagawa-Sakata (PMNS)
lepton flavor matrix \cite{PMNS1,PMNS2} $\widetilde{U}$ defined in matter, we are going to
explore the properties of matter-corrected neutrino mixing and
CP violation in the following two aspects.
\begin{itemize}
\item     We derive new and exact formulas for moduli of the nine
elements of $\widetilde{U}$ and the sides of its three Dirac unitarity triangles
in the complex plane. Different from the previous formulas
of this kind \cite{Xing:2000gg,Xing:2003ez,Zhang:2004hf,Xing:2005gk},
our present results are more symmetric and independent
of the uneasy terms $\widetilde{m}^2_j - m^2_i$ with $m^{}_i$ and
$\widetilde{m}^{}_j$ standing respectively for the genuine neutrino masses
in vacuum and their effective counterparts in matter (for $i, j =1, 2, 3$).
This improvement makes sense because only $\widetilde{\Delta}^{}_{ji}$ are
physical for neutrino oscillations in matter.

\item     We analytically unravel the asymptotic behaviors of
$|\widetilde{U}^{}_{\alpha i}|^2$ and $\widetilde{\Delta}^{}_{ji}$
(for $\alpha = e, \mu, \tau$ and $i,j =1,2,3$) in very dense matter (i.e., when the
matter parameter $A = 2\sqrt{2} ~ G^{}_{\rm F} N^{}_e E$ is considerably large
and even allowed to approach infinity). This is the first time that a full and
analytical understanding of these matter-corrected quantities in the $A\to \infty$
limit has been achieved purely in terms of the fundamental quantities $
|U^{}_{\alpha i}|^2$ and $\Delta^{}_{ji} \equiv m^2_j - m^2_i$,
although their asymptotic behaviors were partly
observed in some previous numerical calculations
(see, e.g., Refs. \cite{Xing:2018lob,Xing:2003ez})
\footnote{In Ref. \cite{Blennow:2004js} Blennow and Ohlsson have
discussed an interesting scenario of the effective two-flavor neutrino mixing 
in the $A \to \infty$ limit and its validity to describe neutrino oscillations 
in a medium with a large but finite electron number density by taking the standard 
parametrization of the effective PMNS matrix $\widetilde{U}$. Our formulas and
main results in the present work are essentially different from theirs.}.
\end{itemize}
Of course, the sum rules that we have derived can also be used to calculate
the effective Jarlskog invariant of CP violation $\widetilde{\cal J}$
\cite{Jarlskog:1985ht} in matter, from which it is straightforward to establish
the Naumov relation between $\widetilde{\cal J}$ and its counterpart
${\cal J}$ in vacuum \cite{Naumov:1991ju}.

It is also worth stressing that our analytical results are parametrization-independent,
and thus they can be used to clarify the confusion associated with the asymptotic
results of $\widetilde{\theta}^{}_{12}$, $\widetilde{\theta}^{}_{13}$,
$\widetilde{\theta}^{}_{23}$ and $\widetilde{\delta}$ in the standard parametrization
of $\widetilde{U}$. The point is that only one degree of freedom is needed to
describe the effective PMNS matrix $\widetilde{U}$ in the $A \to \infty$ limit,
simply because $|\widetilde{U}^{}_{e i}|^2 =1$ and $|\widetilde{U}^{}_{e j}|^2 =
|\widetilde{U}^{}_{e k}|^2 =0$ (for $i \neq j \neq k =1,2,3$) hold in this special
case. So it is always possible to remove $\widetilde{\delta}$ from
$\widetilde{U}$ if the $A \to \infty$ limit is taken, and then we are left with
a trivial flavor mixing angle (e.g., $\widetilde{\theta}^{}_{13} =\pi/2$) and
a nontrivial flavor mixing angle which is neither $\widetilde{\theta}^{}_{12}$ nor
$\widetilde{\theta}^{}_{23}$. This kind of subtle parameter redundancy was not
noticed in the previous papers (see, e.g., Refs. \cite{Xing:2018lob,Wang:2019yfp}),
where specific but misleading values of $\widetilde{\delta}$, $\widetilde{\theta}^{}_{12}$
and $\widetilde{\theta}^{}_{23}$ have been obtained in the $A \to \infty$ limit.

\section{Exact formulas}

In the standard three-flavor scheme, the effective Hamiltonian responsible for
a neutrino beam propagating in a medium can be expressed as
\begin{eqnarray}
{\cal H}^{}_{\rm m} = \frac{1}{2E} U \left(\begin{matrix} m^2_1 & 0 & 0 \\ 0
& m^2_2 & 0 \\ 0 & 0 & m^2_3 \end{matrix}\right) U^\dagger +
\left(\begin{matrix} V^{}_{\rm cc} + V^{}_{\rm nc} & 0 & 0 \\ 0 &
V^{}_{\rm nc} & 0 \\
0 & 0 & V^{}_{\rm nc} \end{matrix}\right)
\equiv \frac{1}{2E}\widetilde U \left(\begin{matrix} \widetilde{m}^2_1 & 0 & 0 \\ 0
& \widetilde{m}^2_2 & 0 \\ 0 & 0 & \widetilde{m}^2_3 \end{matrix}\right)
\widetilde U^\dagger \; ,
\end{eqnarray}
where $V^{}_{\rm cc} = \sqrt{2} \ G^{}_{\rm F} N^{}_e$ and
$V^{}_{\rm nc} = -G^{}_{\rm F} N^{}_n /\sqrt{2}$ are the so-called matter potential
terms arising respectively from weak charged- and neutral-current interactions of
neutrinos with electrons and neutrons in this medium \cite{Wolfenstein:1977ue}.
When an antineutrino beam is concerned, the corresponding effective Hamiltonian
in matter can directly be read off from Eq. (1) with the replacements
$U \to U^*$, $V^{}_{\rm cc} \to -V^{}_{\rm cc}$ and $V^{}_{\rm nc}
\to -V^{}_{\rm nc}$.
Because neutrino (or antineutrino) oscillations depend only on the neutrino
mass-squared differences, it is more convenient to rewrite Eq. (1) in the
following way:
\begin{eqnarray}
{\cal H}^{\prime}_{\rm m} = \frac{1}{2E} \left[U \left(\begin{matrix} 0 & 0 & 0 \\ 0
& \Delta^{}_{21} & 0 \\ 0 & 0 & \Delta^{}_{31} \end{matrix}\right) U^\dagger +
\left(\begin{matrix} A & 0 & 0 \\ 0 & 0 & 0 \\ 0 & 0 & 0 \end{matrix}\right)\right]
\equiv \frac{1}{2E} \left[\widetilde{U} \left(\begin{matrix} 0 & 0 & 0 \\ 0
& \widetilde{\Delta}^{}_{21} & 0 \\ 0 & 0 & \widetilde{\Delta}^{}_{31}
\end{matrix}\right) \widetilde{U}^\dagger + BI\right] \; ,
\end{eqnarray}
where $A = 2 E V^{}_{\rm cc}$ and $B = \widetilde{m}^2_1 - m^2_1 -  2E V^{}_{\rm nc}$,
and $I$ denotes the identity matrix. Given the analytical expressions
of $\widetilde{m}^2_i$ (for $i=1,2,3$) which have been derived in Refs.   \cite{Xing:2000gg,Barger:1980tf,Zaglauer:1988gz}, it is straightforward for us to
obtain
\begin{eqnarray}
&& \widetilde{\Delta}^{}_{21} =
\frac{2}{3} \sqrt{x^2 -3 y} \sqrt{3 \left(1 - z^2\right)} \;\; ,
\nonumber \\
&& \widetilde{\Delta}^{}_{31} =
\frac{1}{3} \sqrt{x^2 -3 y} \left[3z + \sqrt{3 \left(1 - z^2\right)}
\right] \; ,
\nonumber \\
&& B = \frac{1}{3} x - \frac{1}{3} \sqrt{x^2 -3 y} \left[z + \sqrt{3
	\left(1 - z^2\right)} \right] \; \hspace{1cm}
\end{eqnarray}
if three neutrinos have a normal mass ordering (NMO) with $m^{}_1 < m^{}_2 < m^{}_3$
or $\Delta^{}_{31} >0$; or
\begin{eqnarray}
&& \widetilde{\Delta}^{}_{21} =
\frac{1}{3} \sqrt{x^2 - 3 y} \left[3z - \sqrt{3 \left(1 - z^2\right)}\right] \; ,
\nonumber \\
&& \widetilde{\Delta}^{}_{31} = -\frac{2}{3} \sqrt{x^2 - 3 y} \sqrt{3 \left(1 -
	z^2\right)} \;\; ,
\nonumber \\
&& B = \frac{1}{3} x - \frac{1}{3} \sqrt{x^2 -3 y} \left[z - \sqrt{3
	\left(1 - z^2\right)} \right] \; \hspace{1cm}
\end{eqnarray}
if three neutrinos have an inverted mass ordering (IMO) with
$m^{}_3 < m^{}_1 < m^{}_2$ or $\Delta^{}_{31} <0$, where $x$, $y$
and $z$ are given by
\begin{eqnarray}
x & = & \Delta^{}_{21} + \Delta^{}_{31} + A \; ,
\nonumber \\
y & = & \Delta^{}_{21} \Delta^{}_{31} + A \left[\Delta^{}_{21} \left(1 -
|U^{}_{e2}|^2\right) + \Delta^{}_{31} \left(1- |U^{}_{e3}|^2\right)\right] \; ,
\nonumber \\
z & = & \cos\left[\frac{1}{3}\arccos\frac{2 x^3 - 9 xy + 27 A \Delta^{}_{21}
\Delta^{}_{31} |U^{}_{e1}|^2}{2 \sqrt{\left(x^2 - 3y\right)^3}}\right] \; .
\end{eqnarray}
Taking the trace of ${\cal H}^\prime_{\rm m}$ in Eq. (2), we immediately arrive at
\begin{eqnarray}
B = \frac{1}{3} \left(\Delta_{21}^{} + \Delta_{31}^{} + A -
\widetilde \Delta_{21}^{} - \widetilde \Delta_{31}^{}\right) \; .
\end{eqnarray}
Note again that Eqs. (2)---(6) are only valid for neutrino mixing and flavor
oscillations in matter. When an antineutrino beam travelling in a medium is
taken into account, one should make the replacements $U \to U^*$ and
$A \to -A$ for Eqs. (2)---(6).

Eq. (2) allows us to obtain the following sum rules in an easy way:
\begin{eqnarray}
\sum_{i=1}^{3} \widetilde {U}_{\alpha i}^{}  \widetilde {U}_{\beta i}^{*}
\widetilde{\Delta}_{i1}^{} = \sum_{i=1}^{3} {U}_{\alpha i}^{}
{U}_{\beta i}^{*} {\Delta}_{i1}^{} + A \delta_{e \alpha}^{} \delta_{e \beta}^{}
- B \delta_{\alpha\beta} \;,
\end{eqnarray}	
where the Greek and Latin subscripts run over $(e, \mu, \tau)$ and
$(1,2,3)$, respectively. On the other hand, a direct calculation of
${\cal H}_{\rm m}^{\prime 2}$ leads us to another set of sum rules:
\begin{eqnarray}
\sum_{i=1}^{3} \widetilde {U}_{\alpha i}^{}  \widetilde {U}_{\beta i}^{*}
\widetilde{\Delta}_{i1}^{} (\widetilde{\Delta}_{i1}^{} + 2 B)
= \sum_{i=1}^{3} {U}_{\alpha i}^{}   {U}_{\beta i}^{*} {\Delta}_{i1}^{}
\left[{\Delta}_{i1}^{} + A (\delta_{e \alpha}^{} + \delta_{e \beta}^{})
\right]+ A^2 \delta_{e \alpha}^{} \delta_{e \beta}^{}
- B^{2} \delta_{\alpha\beta}^{} \;.
\end{eqnarray}
Eqs. (7) and (8), together with the unitarity conditions of
$U$ and $\widetilde U$,
\begin{eqnarray}
\sum_{i=1}^{3} \widetilde{U}_{\alpha i}^{} \widetilde{U}_{\beta i}^{*}
=\sum_{i=1}^{3} {U}_{\alpha i}^{} {U}_{\beta i}^{*}
=\delta_{\alpha\beta}^{}\;,
\end{eqnarray}
constitute a full set of linear equations of three unknown variables
$\widetilde{U}_{\alpha 1}^{}\widetilde{U}_{\beta 1}^{*}$,
$\widetilde{U}_{\alpha 2}^{}\widetilde{U}_{\beta 2}^{*}$ and
$\widetilde{U}_{\alpha 3}^{}\widetilde{U}_{\beta 3}^{*}$
for two given flavors $\alpha$ and $\beta$. One may therefore solve these
equations and then express $\widetilde{U}_{\alpha i}^{}\widetilde{U}_{\beta i}^{*}$
in terms of $U_{\alpha i}^{} U_{\beta i}^{*}$, $\Delta^{}_{ji}$,
$\widetilde{\Delta}^{}_{ji}$, $A$ and $B$.

\subsection{Moduli of the matrix elements $\widetilde{U}^{}_{\alpha i}$}

Taking $\alpha=\beta$, we obtain a full set of linear
equations of $|\widetilde U_{\alpha i}^{}|^2$ from Eqs. (7)---(9) as follows:
\begin{eqnarray}
&& |\widetilde U_{\alpha 1}^{}|^2 + |\widetilde U_{\alpha 2}^{}|^2
+|\widetilde U_{\alpha 3}^{}|^2 = 1 \; ,
\nonumber \\
&& \widetilde{\Delta}_{21}^{} |\widetilde U_{\alpha 2}^{}|^2
+ \widetilde{\Delta}_{31}^{} |\widetilde U_{\alpha 3}^{}|^2 = \xi \; ,
\nonumber \\
&& \widetilde{\Delta}_{21}^{} (\widetilde{\Delta}_{21}^{} + 2B)
|\widetilde U_{\alpha 2}^{}|^2 + \widetilde{\Delta}_{31}^{}
(\widetilde{\Delta}_{31}^{} + 2B) |\widetilde U_{\alpha 3}^{}|^2
= \zeta \; , \hspace{1cm}
\end{eqnarray}
where
\begin{eqnarray}
\xi & = & \Delta_{21}^{} |{U}_{\alpha 2}^{}|^2 +
\Delta_{31}^{} |{U}_{\alpha 3}^{}|^2+ A \delta_{e \alpha}^{} - B \; ,
\nonumber \\
\zeta & =&  {\Delta}_{21}^{} ({\Delta}_{21}^{} + 2 A \delta_{e
\alpha}^{}) |{U}_{\alpha 2}^{}|^2 + {\Delta}_{31}^{} ({\Delta}_{31}^{} +
2 A \delta_{e \alpha}^{}) |{U}_{\alpha 3}^{}|^2 + A^2 \delta_{e \alpha}^{}
- B^2 \; .
\end{eqnarray}
The solutions of Eq. (10) turn out to be
\begin{align*}
|\widetilde U_{\alpha 1}^{}|^2 & =  \frac{ \zeta - 2 \xi B
- \xi \widetilde{\Delta}_{21}^{} - \xi \widetilde{\Delta}_{31}^{}
+ \widetilde{\Delta}_{21}^{} \widetilde{\Delta}_{31}^{}}
{\widetilde{\Delta}_{21}^{} \widetilde{\Delta}_{31}^{}} \; ,
 \\
|\widetilde U_{\alpha 2}^{}|^2 & =
\frac{\xi\widetilde{ \Delta}_{31}^{} + 2 \xi B - \zeta}
{\widetilde{\Delta}_{21}^{} \widetilde{\Delta}_{32}^{}} \; ,\\
|\widetilde U_{\alpha 3}^{}|^2 & =  \frac{\zeta
- 2 \xi B - \xi \widetilde{\Delta}_{21}^{}}
{\widetilde{\Delta}_{31}^{} \widetilde{\Delta}_{32}^{}} \; ,
\tag{12}
\end{align*}
where $\alpha = e, \mu, \tau$.
With the help of Eqs. (6), (11) and (12), nine $|\widetilde{U}_{\alpha i}^{}|^2$
can be explicitly expressed as
\begin{align*}
|\widetilde{U}_{e 1}^{}|^2  &=  \frac{1}{9} \left[
\frac{\widetilde\Delta_{21}^{} + \widetilde\Delta_{31}^{} +
\Delta_{31}^{}+\Delta_{32}^{} - A}{\widetilde\Delta_{31}^{}} \cdot
\frac{\widetilde\Delta_{21}^{} + \widetilde\Delta_{31}^{} +
\Delta_{21}^{} - \Delta_{32}^{} - A}{\widetilde\Delta_{21}^{} }
\ |U^{}_{e1}|^2 \right.
\nonumber\\
&\hspace{0.48cm}   + \left. \frac{\widetilde\Delta_{21}^{} + \widetilde\Delta_{31}^{}
+ \Delta_{31}^{}+\Delta_{32}^{} - A}{\widetilde\Delta_{31}^{}}
\cdot \frac{\widetilde\Delta_{21}^{} +
\widetilde\Delta_{31}^{} - \Delta_{21}^{} - \Delta_{31}^{} -
A}{\widetilde\Delta_{21}^{}} \ |{U}_{e2}^{}|^2 \right.
\nonumber\\
&\hspace{0.48cm} + \left. \frac{\widetilde\Delta_{21}^{} +
\widetilde\Delta_{31}^{}
+ \Delta_{21}^{} - \Delta_{32}^{} - A}{\widetilde\Delta_{31}^{}}
\cdot \frac{\widetilde\Delta_{21}^{} + \widetilde\Delta_{31}^{}
- \Delta_{21}^{} - \Delta_{31}^{} - A}{\widetilde\Delta_{21}^{}} \
|{U}_{e3}^{}|^2 \right] \; ,
\tag{13a}
\end{align*}
\vspace{-0.6 cm}
\begin{align*}
|\widetilde{U}_{e 2}^{}|^2  & = \frac{1}{9} \left[
\frac{\widetilde\Delta_{32}^{} - \widetilde\Delta_{21}^{} +
\Delta_{31}^{} + \Delta_{32}^{} - A}{\widetilde\Delta_{32}^{}}
\cdot \frac{\widetilde\Delta_{21}^{} -
\widetilde\Delta_{32}^{} + \Delta_{32}^{} - \Delta_{21}^{} +
A}{\widetilde\Delta_{21}^{}} \ |{U}_{e 1}^{}|^2 \right.
\nonumber \\
&\hspace{0.48cm}  + \left. \frac{\widetilde\Delta_{32}^{} - \widetilde\Delta_{21}^{}
+ \Delta_{31}^{}+\Delta_{32}^{} - A}{\widetilde\Delta_{32}^{}} \cdot
\frac{\widetilde\Delta_{21}^{} - \widetilde\Delta_{32}^{} + \Delta_{21}^{}
+ \Delta_{31}^{} + A}{\widetilde\Delta_{21}^{}} \ |{U}_{e 2}^{}|^2 \right.
\nonumber \\
&\hspace{0.48cm}  + \left. \frac{\widetilde\Delta_{32}^{} - \widetilde\Delta_{21}^{}
- \Delta_{32}^{} + \Delta_{21}^{} - A}{\widetilde\Delta_{32}^{}}
\cdot \frac{\widetilde\Delta_{21}^{}
- \widetilde\Delta_{32}^{} + \Delta_{21}^{} + \Delta_{31}^{} +
A}{\widetilde\Delta_{21}^{}} \ |{U}_{e3}^{}|^2 \right] \; ,
\tag{13b}
\end{align*}
\vspace{-0.6 cm}
\begin{align*}
|\widetilde{U}_{e 3}^{}|^2   &=  \frac{1}{9}\left[
\frac{\widetilde\Delta_{31}^{} + \widetilde\Delta_{32}^{} -
\Delta_{31}^{} - \Delta_{32}^{} + A}{\widetilde\Delta_{31}^{}}
\cdot \frac{\widetilde\Delta_{31}^{} +
\widetilde\Delta_{32}^{} + \Delta_{32}^{} - \Delta_{21}^{} +
A}{\widetilde\Delta_{32}^{}} \ |{U}_{e 1}^{}|^2 \right.
\nonumber\\
&\hspace{0.48cm}  + \left. \frac{\widetilde\Delta_{31}^{} + \widetilde\Delta_{32}^{}
- \Delta_{31}^{} - \Delta_{32}^{} + A}{\widetilde\Delta_{31}^{}}
\cdot \frac{\widetilde\Delta_{31}^{}
+ \widetilde\Delta_{32}^{} + \Delta_{21}^{} + \Delta_{31}^{} +
A}{\widetilde\Delta_{32}^{}} \ |{U}_{e2}^{}|^2 \right.
\nonumber\\
&\hspace{0.48cm}  + \left. \frac{\widetilde\Delta_{31}^{} + \widetilde\Delta_{32}^{}
+ \Delta_{21}^{} + \Delta_{31}^{} + A}{\widetilde\Delta_{31}^{}}
\cdot \frac{\widetilde\Delta_{31}^{}
+ \widetilde\Delta_{32}^{} + \Delta_{32}^{} - \Delta_{21}^{} +
A}{\widetilde\Delta_{32}^{}} \ |{U}_{e3}^{}|^2 \right] \; ; \tag{13c}
\end{align*}
and
\begin{align*}
|\widetilde{U}_{\mu 1}^{}|^2  & =  \frac{1}{9}\left[
\frac{\widetilde\Delta_{21}^{} - \widetilde\Delta_{32}^{} +
\Delta_{21}^{}+\Delta_{31}^{} + A}{\widetilde\Delta_{21}^{}}
\cdot \frac{\widetilde\Delta_{31}^{} +
\widetilde\Delta_{32}^{} + \Delta_{21}^{} + \Delta_{31}^{} +
A}{\widetilde\Delta_{31}^{} } \ |{U}_{\mu 1}^{}|^2 \right.
\nonumber\\
&\hspace{0.48cm}  + \left. \frac{\widetilde\Delta_{21}^{} - \widetilde\Delta_{32}^{}
+ \Delta_{32}^{}- \Delta_{21}^{} + A}{\widetilde\Delta_{21}^{}}
\cdot \frac{\widetilde\Delta_{31}^{} +
\widetilde\Delta_{32}^{} + \Delta_{32}^{} - \Delta_{21}^{} +
A}{\widetilde\Delta_{31}^{} } \ |{U}_{\mu 2}^{}|^2 \right.
\nonumber\\
&\hspace{0.48cm}  + \left. \frac{\widetilde\Delta_{21}^{} - \widetilde\Delta_{32}^{}
- \Delta_{31}^{} - \Delta_{32}^{} + A}{\widetilde\Delta_{21}^{}}
\cdot \frac{\widetilde\Delta_{31}^{}
+ \widetilde\Delta_{32}^{} - \Delta_{31}^{} - \Delta_{32}^{} +
A}{\widetilde\Delta_{31}^{} } \ |{U}_{\mu 3}^{}|^2 \right] \; ,
\tag{14a}
\end{align*}
\begin{align*}
|\widetilde{U}_{\mu 2}^{}|^2  & =  \frac{1}{9}\left[
\frac{\widetilde\Delta_{21}^{} + \widetilde\Delta_{31}^{} -
\Delta_{21}^{} - \Delta_{31}^{} - A}{\widetilde\Delta_{21}^{}}
\cdot \frac{\widetilde\Delta_{31}^{} +
\widetilde\Delta_{32}^{} + \Delta_{21}^{} + \Delta_{31}^{} +
A}{\widetilde\Delta_{32}^{} } \ |{U}_{\mu 1}^{}|^2 \right.
\nonumber\\
&\hspace{0.48cm}  + \left. \frac{\widetilde\Delta_{21}^{} + \widetilde\Delta_{31}^{}
+ \Delta_{21}^{} -  \Delta_{32}^{} - A}{\widetilde\Delta_{21}^{}}
\cdot \frac{\widetilde\Delta_{31}^{}
+ \widetilde\Delta_{32}^{} + \Delta_{32}^{} - \Delta_{21}^{} +
A}{\widetilde\Delta_{32}^{} } \ |{U}_{\mu 2}^{}|^2 \right.
\nonumber\\
&\hspace{0.48cm} + \left. \frac{\widetilde\Delta_{21}^{} + \widetilde\Delta_{31}^{}
+ \Delta_{31}^{} + \Delta_{32}^{} - A}{\widetilde\Delta_{21}^{}}
\cdot \frac{\widetilde\Delta_{31}^{}
+ \widetilde\Delta_{32}^{} - \Delta_{31}^{} - \Delta_{32}^{} +
A}{\widetilde\Delta_{32}^{} } \ |{U}_{\mu 3}^{}|^2 \right] \; ,
\tag{14b}
\end{align*}
\vspace{-0.6 cm}
\begin{align*}
|\widetilde{U}_{\mu 3}^{}|^2  & =  \frac{1}{9}\left[
\frac{\widetilde\Delta_{21}^{} + \widetilde\Delta_{31}^{} -
\Delta_{21}^{} - \Delta_{31}^{} - A}{\widetilde\Delta_{31}^{}}
\cdot \frac{\widetilde\Delta_{32}^{} -
\widetilde\Delta_{21}^{} - \Delta_{21}^{} - \Delta_{31}^{} -
A}{\widetilde\Delta_{32}^{} } \ |{U}_{\mu 1}^{}|^2 \right.
\nonumber\\
&\hspace{0.48cm}  + \left. \frac{\widetilde\Delta_{21}^{} + \widetilde\Delta_{31}^{}
+ \Delta_{21}^{} - \Delta_{32}^{} - A}{\widetilde\Delta_{31}^{}}
\cdot \frac{\widetilde\Delta_{32}^{}
- \widetilde\Delta_{21}^{} - \Delta_{32}^{} + \Delta_{21}^{} -
A}{\widetilde\Delta_{32}^{} } \ |{U}_{\mu 2}^{}|^2 \right.
\nonumber\\
&\hspace{0.48cm}  + \left. \frac{\widetilde\Delta_{21}^{} + \widetilde\Delta_{31}^{}
+ \Delta_{31}^{} + \Delta_{32}^{} - A}{\widetilde\Delta_{31}^{}}
\cdot \frac{\widetilde\Delta_{32}^{}
- \widetilde\Delta_{21}^{} + \Delta_{31}^{} + \Delta_{32}^{} -
A}{\widetilde\Delta_{32}^{} } \ |{U}_{\mu 3}^{}|^2 \right] \; ;
\tag{14c}
\end{align*}	
as well as
\begin{align*}
|\widetilde{U}_{\tau 1}^{}|^2  & =  \frac{1}{9}\left[
\frac{\widetilde\Delta_{21}^{} - \widetilde\Delta_{32}^{} +
\Delta_{21}^{}+\Delta_{31}^{} + A}{\widetilde\Delta_{21}^{}}
\cdot \frac{\widetilde\Delta_{31}^{} +
\widetilde\Delta_{32}^{} + \Delta_{21}^{} + \Delta_{31}^{} +
A}{\widetilde\Delta_{31}^{} } \ |{U}_{\tau 1}^{}|^2 \right.
\nonumber\\
&\hspace{0.48cm}  + \left. \frac{\widetilde\Delta_{21}^{} - \widetilde\Delta_{32}^{}
+ \Delta_{32}^{}- \Delta_{21}^{} + A}{\widetilde\Delta_{21}^{}}
\cdot \frac{\widetilde\Delta_{31}^{} +
\widetilde\Delta_{32}^{} + \Delta_{32}^{} - \Delta_{21}^{} +
A}{\widetilde\Delta_{31}^{} } \ |{U}_{\tau 2}^{}|^2 \right.
\nonumber\\
&\hspace{0.48cm}  + \left. \frac{\widetilde\Delta_{21}^{} - \widetilde\Delta_{32}^{}
- \Delta_{31}^{} - \Delta_{32}^{} + A}{\widetilde\Delta_{21}^{}}
\cdot \frac{\widetilde\Delta_{31}^{}
+ \widetilde\Delta_{32}^{} - \Delta_{31}^{} - \Delta_{32}^{} +
A}{\widetilde\Delta_{31}^{} } \ |{U}_{\tau 3}^{}|^2
\right] \; , \tag{15a}
\end{align*}
\vspace{-0.6 cm}
\begin{align*}
|\widetilde{U}_{\tau 2}^{}|^2  & =  \frac{1}{9}\left[
\frac{\widetilde\Delta_{21}^{} + \widetilde\Delta_{31}^{} -
\Delta_{21}^{} - \Delta_{31}^{} - A}{\widetilde\Delta_{21}^{}}
\cdot \frac{\widetilde\Delta_{31}^{} +
\widetilde\Delta_{32}^{} + \Delta_{21}^{} + \Delta_{31}^{} +
A}{\widetilde\Delta_{32}^{} } \ |{U}_{\tau 1}^{}|^2 \right.
\nonumber\\
&\hspace{0.48cm}  + \left. \frac{\widetilde\Delta_{21}^{} + \widetilde\Delta_{31}^{}
+ \Delta_{21}^{} -  \Delta_{32}^{} - A}{\widetilde\Delta_{21}^{}}
\cdot \frac{\widetilde\Delta_{31}^{}
+ \widetilde\Delta_{32}^{} + \Delta_{32}^{} - \Delta_{21}^{} +
A}{\widetilde\Delta_{32}^{} } \ |{U}_{\tau 2}^{}|^2 \right.
\nonumber\\
&\hspace{0.48cm}  + \left. \frac{\widetilde\Delta_{21}^{} + \widetilde\Delta_{31}^{}
+ \Delta_{31}^{} + \Delta_{32}^{} - A}{\widetilde\Delta_{21}^{}}
\cdot \frac{\widetilde\Delta_{31}^{}
+ \widetilde\Delta_{32}^{} - \Delta_{31}^{} - \Delta_{32}^{} +
A}{\widetilde\Delta_{32}^{} } \ |{U}_{\tau 3}^{}|^2 \right] \; ,
\tag{15b}
\end{align*}
\vspace{-0.6 cm}
\begin{align*}
|\widetilde{U}_{\tau 3}^{}|^2  & =  \frac{1}{9}\left[
\frac{\widetilde\Delta_{21}^{} + \widetilde\Delta_{31}^{} -
\Delta_{21}^{} - \Delta_{31}^{} - A}{\widetilde\Delta_{31}^{}}
\cdot \frac{\widetilde\Delta_{32}^{} -
\widetilde\Delta_{21}^{} - \Delta_{21}^{} - \Delta_{31}^{} -
A}{\widetilde\Delta_{32}^{} } \ |{U}_{\tau 1}^{}|^2 \right.
\nonumber\\
&\hspace{0.48cm}  + \left. \frac{\widetilde\Delta_{21}^{} + \widetilde\Delta_{31}^{}
+ \Delta_{21}^{} - \Delta_{32}^{} - A}{\widetilde\Delta_{31}^{}}
\cdot \frac{\widetilde\Delta_{32}^{}
- \widetilde\Delta_{21}^{} - \Delta_{32}^{} + \Delta_{21}^{} -
A}{\widetilde\Delta_{32}^{} } \ |{U}_{\tau 2}^{}|^2 \right.
\nonumber\\
&\hspace{0.48cm}  + \left. \frac{\widetilde\Delta_{21}^{} + \widetilde\Delta_{31}^{}
+ \Delta_{31}^{} + \Delta_{32}^{} - A}{\widetilde\Delta_{31}^{}}
\cdot \frac{\widetilde\Delta_{32}^{}
- \widetilde\Delta_{21}^{} + \Delta_{31}^{} + \Delta_{32}^{} -
A}{\widetilde\Delta_{32}^{} } \ |{U}_{\tau 3}^{}|^2 \right] \; .
\tag{15c}
\end{align*}	
Since the expressions of $\widetilde{\Delta}^{}_{ji}$ have been given
in Eq. (3) for the NMO case and in Eq. (4) for the IMO case, it is
straightforward to calculate $|\widetilde{U}^{}_{\alpha i}|^2$ by
taking a specific value of the matter parameter $A$ and inputting
the experimental values of two neutrino mass-squared differences
and four flavor mixing parameters in vacuum.

From a phenomenological point of view, we emphasize that the analytical
results of $|\widetilde{U}^{}_{\alpha i}|^2$ obtained above are more
advantageous than the previous ones obtained in Refs.
\cite{Zhang:2004hf,Xing:2005gk} in two aspects. First, the present
results are more symmetric and transparent in reflecting the relations
between $|\widetilde{U}^{}_{\alpha i}|^2$ and $|{U}^{}_{\alpha i}|^2$.
Second, the present expressions of $|\widetilde{U}^{}_{\alpha i}|^2$
are free from the uneasy terms $\widetilde{m}^2_j - m^2_i$ in which
the effective neutrino masses $\widetilde{m}^2_j$ do not have a definite
physical meaning. In fact, only $\widetilde{\Delta}^{}_{ji}$ are physical
for neutrino oscillations in matter.

Once again, one should keep in mind that the above results are only
valid for a neutrino beam travelling in matter. It is necessary to
make the replacements $U \to U^*$ and $A \to -A$ when an antineutrino
beam is taken into account.

\subsection{Sides of the Dirac unitarity triangles}

As in vacuum, the orthogonality conditions of $\widetilde{U}$ given
in Eq. (9) can define three distinct unitarity triangles in the
complex plane --- the so-called {\it effective} Dirac unitarity triangles
in matter \cite{Xing:2016ymg,Xing:2015wzz}
\begin{align*}
\widetilde{\triangle}^{}_e : & \hspace{0.2cm}
\widetilde{U}^{}_{\mu 1} \widetilde{U}^*_{\tau 1} +
\widetilde{U}^{}_{\mu 2} \widetilde{U}^*_{\tau 2} +
\widetilde{U}^{}_{\mu 3} \widetilde{U}^*_{\tau 3} = 0 \; , \hspace{0.8cm}
\nonumber \\
\widetilde{\triangle}^{}_\mu : & \hspace{0.2cm}
\widetilde{U}^{}_{\tau 1} \widetilde{U}^*_{e 1} +
\widetilde{U}^{}_{\tau 2} \widetilde{U}^*_{e 2} +
\widetilde{U}^{}_{\tau 3} \widetilde{U}^*_{e 3} = 0 \; ,
\nonumber \\
\widetilde{\triangle}^{}_\tau : & \hspace{0.2cm}
\widetilde{U}^{}_{e 1} \widetilde{U}^*_{\mu 1} +
\widetilde{U}^{}_{e 2} \widetilde{U}^*_{\mu 2} +
\widetilde{U}^{}_{e 3} \widetilde{U}^*_{\mu 3} = 0 \; ,
\tag{16}
\end{align*}
which are insensitive to a redefinition of the phases of three neutrino
fields and thus have nothing to do with the Majorana phases of the PMNS matrix $U$.
Each of these three triangle is named after the flavor index that does not
show up in its three sides. The areas of $\widetilde{\triangle}^{}_e$,
$\widetilde{\triangle}^{}_\mu$ and $\widetilde{\triangle}^{}_\tau$
are all equal to half of the magnitude of the {\it effective} Jarlskog invariant
of CP violation in matter, denoted by $\widetilde{\cal J}$. The latter, together
with its fundamental counterpart ${\cal J}$
in vacuum \cite{Jarlskog:1985ht}, is defined as
\begin{align*}
{\rm Im} (U^{}_{\alpha i} U^{}_{\beta j} U^*_{\alpha j} U^*_{\beta i})
& =  {\cal J} \sum_\gamma \varepsilon^{}_{\alpha \beta \gamma}
\sum_k \varepsilon^{}_{ijk} \; , \hspace{0.8cm}
\nonumber \\
{\rm Im} (\widetilde{U}^{}_{\alpha i} \widetilde{U}^{}_{\beta j}
\widetilde{U}^*_{\alpha j} \widetilde{U}^*_{\beta i})
& =  \widetilde{\cal J} \sum_\gamma \varepsilon^{}_{\alpha \beta \gamma}
\sum_k \varepsilon^{}_{ijk} \; ,
\tag{17}
\end{align*}
where $\varepsilon^{}_{\alpha \beta \gamma}$ and $\varepsilon^{}_{ijk}$ are the
three-dimension Levi-Civita symbols.

Taking $\alpha \neq \beta$, Eqs. (7)---(9) can now be simplified to the following
set of linear equations of three variables
$\widetilde U_{\alpha i}^{} \widetilde U_{\beta i}^{*}$ (for $i =1,2,3$):
\begin{align*}
& \widetilde U_{\alpha 1}^{} \widetilde U_{\beta 1}^{*} +
\widetilde U_{\alpha 2}^{} \widetilde U_{\beta
2}^{*} + \widetilde U_{\alpha 3}^{} \widetilde U_{\beta 3}^{*} = 0 \; ,
\nonumber \\
& \widetilde{\Delta}_{21}^{} \widetilde U_{\alpha 2}^{}
\widetilde U_{\beta 2}^{*} + \widetilde{\Delta}_{31}^{} \widetilde
U_{\alpha 3}^{} \widetilde U_{\beta 3}^{*} = \xi^{\prime} \; ,
\nonumber \\
& \widetilde{\Delta}_{21}^{}(\widetilde{\Delta}_{21}^{} + 2B)
\widetilde U_{\alpha 2}^{} \widetilde U_{\beta 2}^{*} +
\widetilde{\Delta}_{31}^{}(\widetilde{\Delta}_{31}^{} + 2B)
\widetilde U_{\alpha 3}^{} \widetilde U_{\beta 3}^{*}
= \zeta^{\prime} \; , \hspace{0.8cm}
\tag{18}
\end{align*}
where
\begin{align*}
\xi^{\prime} & = \Delta_{21}^{}  U_{\alpha 2}^{}
U_{\beta 2}^{*}  + \Delta_{31}^{}  U_{\alpha 3}^{}
U_{\beta 3}^{*} \;,
\nonumber\\
\zeta^{\prime} & =  {\Delta}_{21}^{} \left[{\Delta}_{21}^{} +
A (\delta_{e \alpha}^{} + \delta_{e \beta}^{}) \right]
U_{\alpha 2}^{} U_{\beta 2}^{*} + {\Delta}_{31}^{}
\left[{\Delta}_{31}^{} + A (\delta_{e \alpha}^{}
+ \delta_{e \beta}^{}) \right] U_{\alpha 3}^{} U_{\beta 3}^{*} \;.
\tag{19}
\end{align*}
Solving Eq. (18) in a straightforward way, we are left with the solutions
\begin{align*}
\widetilde U_{\alpha 1}^{} \widetilde U_{\beta 1}^{*} & =  \frac{
\zeta^{\prime} - 2 \xi^{\prime} B - \xi^{\prime}
\widetilde{\Delta}_{21}^{} - \xi^\prime \widetilde{\Delta}_{31}^{}}
{\widetilde{\Delta}_{21}^{} \widetilde{\Delta}_{31}^{}} \; ,
\nonumber\\
\widetilde U_{\alpha 2}^{} \widetilde U_{\beta 2}^{*} & =
\frac{\xi^{\prime}\widetilde{\Delta}_{31}^{}
+ 2 \xi^{\prime} B - \zeta^{\prime} }{\widetilde{\Delta}_{21}^{}
\widetilde{\Delta}_{32}^{}} \; ,
\nonumber\\
\widetilde U_{\alpha 3}^{} \widetilde U_{\beta 3}^{*} & =
\frac{\zeta^{\prime} - 2 \xi^{\prime} B - \xi^{\prime}
\widetilde{\Delta}_{21}^{} }{\widetilde{\Delta}_{31}^{}
\widetilde{\Delta}_{32}^{}} \; . \tag{20}
\end{align*}
After Eq. (6) is taken into account, the explicit expressions of
nine $\widetilde{U}_{\alpha i}^{} \widetilde{U}_{\beta i}^{*}$ can be
obtained from Eqs. (19) and (20). Namely,
\begin{align*}
\widetilde{U}_{\mu 1}^{} \widetilde{U}_{\tau 1}^{*}  & =  \frac{1}{3}\left[
\frac{\widetilde\Delta_{21}^{} + \widetilde\Delta_{31}^{} +
\Delta_{21}^{}-\Delta_{32}^{} + 2 A}{\widetilde\Delta_{21}^{}}
\cdot \frac{\Delta_{31}^{}}{\widetilde\Delta_{31}^{}} \
{U}_{\mu 1}^{} {U}_{\tau 1}^{*} \right.
\nonumber \\
&\hspace{0.48cm} + \left. \frac{\widetilde\Delta_{21}^{} + \widetilde\Delta_{31}^{} -
\Delta_{21}^{}-\Delta_{31}^{} + 2 A}{\widetilde\Delta_{21}^{}}
\cdot \frac{\Delta_{32}^{}}{\widetilde\Delta_{31}^{}} \
{U}_{\mu 2}^{} {U}_{\tau 2}^{*} \right] \; , \tag{21a}
\end{align*}
\vspace{-0.6 cm}
\begin{align*}
\widetilde{U}_{\mu 2}^{} \widetilde{U}_{\tau 2}^{*}  & =  \frac{1}{3}\left[
\frac{\widetilde\Delta_{32}^{} - \widetilde\Delta_{21}^{} +
\Delta_{31}^{} + \Delta_{32}^{} + 2 A}{\widetilde\Delta_{32}^{}}
\cdot \frac{\Delta_{21}^{}}{\widetilde\Delta_{21}^{}} \
{U}_{\mu 2}^{} {U}_{\tau 2}^{*} \right.
\nonumber\\
&\hspace{0.48cm} + \left. \frac{\widetilde\Delta_{32}^{} - \widetilde\Delta_{21}^{} +
\Delta_{21}^{} - \Delta_{32}^{} + 2 A}{\widetilde\Delta_{32}^{}}
\cdot \frac{\Delta_{31}^{}}{\widetilde\Delta_{21}^{}} \
{U}_{\mu 3}^{} {U}_{\tau 3}^{*} \right] \; , \tag{21b}
\end{align*}
\vspace{-0.6 cm}
\begin{align*}
\widetilde{U}_{\mu 3}^{} \widetilde{U}_{\tau 3}^{*}  & =  \frac{1}{3}\left[
\frac{\widetilde\Delta_{31}^{} + \widetilde\Delta_{32}^{} +
\Delta_{31}^{} + \Delta_{21}^{} - 2 A}{\widetilde\Delta_{31}^{}}
\cdot \frac{\Delta_{32}^{}}{\widetilde\Delta_{32}^{}} \
{U}_{\mu 3}^{} {U}_{\tau 3}^{*} \right.
\nonumber\\
&\hspace{0.48cm} - \left. \frac{\widetilde\Delta_{31}^{} + \widetilde\Delta_{32}^{} -
\Delta_{31}^{} - \Delta_{32}^{} - 2 A}{\widetilde\Delta_{31}^{}}
\cdot \frac{\Delta_{21}^{}}{\widetilde\Delta_{32}^{}} \
{U}_{\mu 1}^{} {U}_{\tau 1}^{*} \right] \; ; \tag{21c}
\end{align*}	
and
\begin{align*}
\widetilde{U}_{\tau 1}^{} \widetilde{U}_{e 1}^{*}  & =  \frac{1}{3}\left[
\frac{\widetilde\Delta_{21}^{} + \widetilde\Delta_{31}^{} +
\Delta_{21}^{} - \Delta_{32}^{} - A}{\widetilde\Delta_{21}^{}}
\cdot \frac{\Delta_{31}^{}}{\widetilde\Delta_{31}^{}} \
{U}_{\tau 1}^{} {U}_{e 1}^{*} \right.
\nonumber \\
&\hspace{0.48cm} + \left. \frac{\widetilde\Delta_{21}^{} + \widetilde\Delta_{31}^{} -
\Delta_{21}^{}-\Delta_{31}^{} - A}{\widetilde\Delta_{21}^{}}
\cdot \frac{\Delta_{32}^{}}{\widetilde\Delta_{31}^{}} \
{U}_{\tau 2}^{} {U}_{e 2}^{*} \right] \; ,
\tag{22a}
\end{align*}
\vspace{-0.6 cm}
\begin{align*}
\widetilde{U}_{\tau 2}^{} \widetilde{U}_{e 2}^{*}  & =  \frac{1}{3}\left[
\frac{\widetilde\Delta_{32}^{} - \widetilde\Delta_{21}^{} +
\Delta_{31}^{} + \Delta_{32}^{} - A}{\widetilde\Delta_{32}^{}}
\cdot \frac{\Delta_{21}^{}}{\widetilde\Delta_{21}^{}} \
{U}_{\tau 2}^{} {U}_{e 2}^{*} \right.
\nonumber \\
&\hspace{0.48cm} + \left. \frac{\widetilde\Delta_{32}^{} - \widetilde\Delta_{21}^{} +
\Delta_{21}^{} - \Delta_{32}^{} - A}{\widetilde\Delta_{32}^{}}
\cdot \frac{\Delta_{31}^{}}{\widetilde\Delta_{21}^{}} \
{U}_{\tau 3}^{} {U}_{e 3}^{*} \right] \; ,
\tag{22b}
\end{align*}
\vspace{-0.6 cm}
\begin{align*}
\widetilde{U}_{\tau 3}^{} \widetilde{U}_{e 3}^{*}  & =  \frac{1}{3}\left[
\frac{\widetilde\Delta_{31}^{} + \widetilde\Delta_{32}^{} +
\Delta_{31}^{} + \Delta_{21}^{} + A}{\widetilde\Delta_{31}^{}}
\cdot \frac{\Delta_{32}^{}}{\widetilde\Delta_{32}^{}} \
{U}_{\tau 3}^{} {U}_{e 3}^{*} \right.
\nonumber \\
&\hspace{0.48cm} - \left. \frac{\widetilde\Delta_{31}^{} + \widetilde\Delta_{32}^{} -
\Delta_{31}^{} - \Delta_{32}^{} +A}{\widetilde\Delta_{31}^{}}
\cdot \frac{\Delta_{21}^{}}{\widetilde\Delta_{32}^{}} \
{U}_{\tau 1}^{} {U}_{e 1}^{*} \right] \; ;
\tag{22c}
\end{align*}		
as well as
\begin{align*}
\widetilde{U}_{e 1}^{} \widetilde{U}_{\mu 1}^{*}  & =  \frac{1}{3}\left[
\frac{\widetilde\Delta_{21}^{} + \widetilde\Delta_{31}^{} +
\Delta_{21}^{} - \Delta_{32}^{} - A}{\widetilde\Delta_{21}^{}}
\cdot \frac{\Delta_{31}^{}}{\widetilde\Delta_{31}^{}} \
{U}_{e 1}^{} {U}_{\mu 1}^{*} \right.
\nonumber \\
&\hspace{0.48cm} + \left. \frac{\widetilde\Delta_{21}^{} + \widetilde\Delta_{31}^{} -
\Delta_{21}^{}-\Delta_{31}^{} - A}{\widetilde\Delta_{21}^{}}
\cdot \frac{\Delta_{32}^{}}{\widetilde\Delta_{31}^{}} \
{U}_{e 2}^{} {U}_{\mu 2}^{*} \right] \; ,
\tag{23a}
\end{align*}
\vspace{-0.6 cm}
\begin{align*}
\widetilde{U}_{e 2}^{} \widetilde{U}_{\mu 2}^{*}  & =  \frac{1}{3}\left[
\frac{\widetilde\Delta_{32}^{} - \widetilde\Delta_{21}^{} +
\Delta_{31}^{} + \Delta_{32}^{} - A}{\widetilde\Delta_{32}^{}}
\cdot \frac{\Delta_{21}^{}}{\widetilde\Delta_{21}^{}} \
{U}_{e 2}^{} {U}_{\mu 2}^{*} \right.
\nonumber \\
&\hspace{0.48cm} + \left. \frac{\widetilde\Delta_{32}^{} - \widetilde\Delta_{21}^{} +
\Delta_{21}^{} - \Delta_{32}^{} - A}{\widetilde\Delta_{32}^{}}
\cdot \frac{\Delta_{31}^{}}{\widetilde\Delta_{21}^{}} \
{U}_{e 3}^{} {U}_{\mu 3}^{*} \right] \; ,
\tag{23b}
\end{align*}
\vspace{-0.6 cm}
\begin{align*}
\widetilde{U}_{e 3}^{} \widetilde{U}_{\mu 3}^{*}  & =  \frac{1}{3}\left[
\frac{\widetilde\Delta_{31}^{} + \widetilde\Delta_{32}^{} +
\Delta_{31}^{} + \Delta_{21}^{} + A}{\widetilde\Delta_{31}^{}}
\cdot \frac{\Delta_{32}^{}}{\widetilde\Delta_{32}^{}} \
{U}_{e 3}^{} {U}_{\mu 3}^{*} \right.
\nonumber \\
&\hspace{0.48cm} - \left. \frac{\widetilde\Delta_{31}^{} + \widetilde\Delta_{32}^{} -
\Delta_{31}^{} - \Delta_{32}^{} + A}{\widetilde\Delta_{31}^{}}
\cdot \frac{\Delta_{21}^{}}{\widetilde\Delta_{32}^{}} \
{U}_{e 1}^{} {U}_{\mu 1}^{*} \right] \; ,
\tag{23c}
\end{align*}		
where $\widetilde{\Delta}^{}_{ji}$ have been given
in Eq. (3) for the NMO case and in Eq. (4) for the IMO case. It is
obvious that the shapes of three Dirac unitarity triangles in vacuum
will be deformed by matter effects, and this implies the change of
$\widetilde{\cal J}$ as compared with ${\cal J}$.

With the help of Eqs. (21)---(23), one may calculate $\widetilde{\cal J}$
by using any two sides of the Dirac unitarity triangle
$\widetilde{\Delta}^{}_\alpha$ (for $\alpha = e, \mu, \tau$). For example,
\begin{align*}
\widetilde{\cal J} & =  {\rm Im}\left[\left(\widetilde{U}^{}_{\mu 2}
\widetilde{U}^*_{\tau 2}\right) \left(\widetilde{U}^{}_{\mu 3}
\widetilde{U}^*_{\tau 3}\right)^*\right]
\nonumber \\
& =  \frac{\Delta^{}_{21} \Delta^{}_{31}}{\widetilde{\Delta}^{}_{21}
\widetilde{\Delta}^{}_{31} \widetilde{\Delta}^2_{32}}
{\rm Im}\left[\left(\widetilde{\Delta}^{}_{31} - \Delta^{}_{21}\right)
\left(\Delta^{}_{31} - \widetilde{\Delta}^{}_{21}\right)
U^{}_{\mu 2} U^{}_{\tau 3} U^*_{\mu 3} U^*_{\tau 2} \right.
\nonumber \\
& \hspace{0.48cm}  + \left.
\left(\widetilde{\Delta}^{}_{31} - \Delta^{}_{31}\right)
\left(\Delta^{}_{21} - \widetilde{\Delta}^{}_{21}\right)
U^{*}_{\mu 2} U^{*}_{\tau 3} U^{}_{\mu 3} U^{}_{\tau 2}\right]
= \frac{\Delta^{}_{21} \Delta^{}_{31} \Delta^{}_{32}}
{\widetilde{\Delta}^{}_{21} \widetilde{\Delta}^{}_{31}
\widetilde{\Delta}^{}_{32}} {\cal J} \; , \hspace{0.6cm}
\tag{24}
\end{align*}
where ${\cal J} = {\rm Im}\left(U^{}_{\mu 2} U^{}_{\tau 3} U^*_{\mu 3}
U^*_{\tau 2}\right)$ has been used. One can see that Eq. (24) is just the
well-known Naumov relation between $\widetilde{\cal J}$ and ${\cal J}$
\cite{Naumov:1991ju}.

The above parametrization-independent expressions of
$|\widetilde{U}^{}_{\alpha i}|^2$ and $\widetilde{U}^{}_{\alpha i}
\widetilde{U}^*_{\beta i}$ can easily be used to derive the effective
neutrino mixing angles ($\widetilde{\theta}^{}_{12}$, $\widetilde{\theta}^{}_{13}$,
$\widetilde{\theta}^{}_{23}$) and the effective CP-violating phase
($\widetilde{\delta}$) in the standard parametrization of $\widetilde{U}$,
\begin{align*}
\widetilde{U} = \left(\begin{matrix}
\widetilde{c}^{}_{12} \widetilde{c}^{}_{13} & \widetilde{s}^{}_{12}
\widetilde{c}^{}_{13} & \widetilde{s}^{}_{13} e^{-{\rm i} \widetilde{\delta}}
\cr -\widetilde{s}^{}_{12} \widetilde{c}^{}_{23} - \widetilde{c}^{}_{12}
\widetilde{s}^{}_{13} \widetilde{s}^{}_{23} e^{{\rm i} \widetilde{\delta}} &
\widetilde{c}^{}_{12} \widetilde{c}^{}_{23} -
\widetilde{s}^{}_{12} \widetilde{s}^{}_{13} \widetilde{s}^{}_{23}
e^{{\rm i} \widetilde{\delta}} & \widetilde{c}^{}_{13}
\widetilde{s}^{}_{23} \cr \widetilde{s}^{}_{12} \widetilde{s}^{}_{23} -
\widetilde{c}^{}_{12} \widetilde{s}^{}_{13} \widetilde{c}^{}_{23}
e^{{\rm i} \widetilde{\delta}} & - \widetilde{c}^{}_{12} \widetilde{s}^{}_{23}
- \widetilde{s}^{}_{12} \widetilde{s}^{}_{13} \widetilde{c}^{}_{23}
e^{{\rm i} \widetilde{\delta}} &  \widetilde{c}^{}_{13} \widetilde{c}^{}_{23} \cr
\end{matrix} \right) \; , \hspace{0.3cm} \tag{25}
\end{align*}
in which $\widetilde{c}^{}_{ij} \equiv \cos\widetilde{\theta}^{}_{ij}$ and
$\widetilde{s}^{}_{ij} \equiv \sin\widetilde{\theta}^{}_{ij}$
with $\widetilde{\theta}^{}_{ij}$ lying in the first quadrant
(for $ij = 12, 13, 23$), and $\widetilde{\delta}$ is allowed to vary between
$0$ and $2\pi$. For instance,
\begin{align*}
\tan\widetilde{\theta}^{}_{12} = \frac{|\widetilde{U}^{}_{e2}|}
{|\widetilde{U}^{}_{e1}|} \; , \quad
\sin\widetilde{\theta}^{}_{13} = |\widetilde{U}^{}_{e3}| \; , \quad
\tan\widetilde{\theta}^{}_{23} = \frac{|\widetilde{U}^{}_{\mu 3}|}
{|\widetilde{U}^{}_{\tau 3}|} \; , \hspace{0.3cm}
\tag{26}
\end{align*}
where the moduli of the effective PMNS matrix elements have been given
in Eqs. (13)---(15), and then $\sin\widetilde{\delta}$ can be obtained from the
Toshev relation $\sin 2\widetilde{\theta}^{}_{23} \sin\widetilde{\delta} =
\sin 2\theta^{}_{23} \sin\delta$ \cite{Toshev:1991ku}.

\section{Asymptotic behaviors}

Now we apply the exact formulas of $|\widetilde{U}^{}_{\alpha i}|^2$ to
an extreme case, in which the matter density is considerably large or equivalent
to $A \to \infty$, to examine the asymptotic behaviors of
$|\widetilde{U}^{}_{\alpha i}|^2$. Although the behaviors of
$|\widetilde{U}^{}_{\alpha i}|^2$ changing with the matter parameter $A$
have been numerically illustrated in the literature (see, e.g., Ref. \cite{Xing:2018lob}),
a comprehensive and analytical understanding of their asymptotic properties
in the $A \to \infty$ limit has been lacking. On the other hand, it has
been shown that in the standard parametrization of $\widetilde{U}$ the
effective CP-violating phase $\widetilde{\delta}$ approaches a finite value
even if $\widetilde{\theta}^{}_{13} \to \pi/2$ in the $A \to \infty$ limit
(see, e.g., Refs. \cite{Xing:2018lob,Wang:2019yfp}). This result is confusing
because $\widetilde{\delta}$ can be rotated away when
$|\widetilde{U}^{}_{e 1}|^2 = |\widetilde{U}^{}_{e 2}|^2 =
|\widetilde{U}^{}_{\mu 3}|^2 = |\widetilde{U}^{}_{\tau 3}|^2 = 0$ holds
as a result of $\cos\widetilde{\theta}^{}_{13} \to 0$.
We are going to clarify this parameter redundancy by demonstrating that
$\widetilde{U}$ only contains one degree of freedom when $A$ approaches
infinity.

\subsection{Case A: (NMO, $\nu$)}

Let us first consider the case of a neutrino beam ($\nu$) propagating in matter
with a normal mass ordering (NMO). When taking $A \to \infty$, we can simplify
Eq. (3) and arrive at
\begin{align*}
& \widetilde{\Delta}^{}_{21} = \sqrt{p^2 - 4q} \;\; ,
\nonumber \\
&\widetilde{\Delta}^{}_{31} = \Delta^{}_{21} + \Delta^{}_{31} + A -
\frac{1}{2}\left(3 p - \sqrt{p^2 - 4q}\right) \; , \hspace{1cm}
\nonumber \\
& B = \frac{1}{2} \left(p - \sqrt{p^2 - 4 q}\right) \; ,
\tag{27}
\end{align*}
where $p = \Delta_{21}^{} \left( 1 - |U_{e2}^{}|^2 \right)
+ \Delta_{31}^{} \left( 1 - |U_{e3}^{}|^2 \right)$ and
$q = \Delta_{21}^{} \Delta_{31}^{} |U_{e1}^{}|^2$. In a good approximation,
we find that $\widetilde \Delta_{21}^{} \simeq \Delta_{31}^{}
\left(1-|U_{e3}^{}|^2\right) - \Delta^{}_{21} |U_{e1}^{}|^2$ is finite
and $\widetilde \Delta_{31}^{} \simeq \widetilde \Delta_{32}^{} \simeq A$
approaches infinity in the $A \to \infty$ limit.

With the help of Eq. (27), one may use Eqs. (13)---(15) to calculate the nine
elements of $\widetilde{U}$ in the $A \to \infty$ limit. The results are
\begin{align*}
|\widetilde{U}_{e1}^{}|^2  & =  |\widetilde{U}_{e2}^{}|^2 =
|\widetilde{U}_{\mu 3}^{}|^2 = |\widetilde{U}_{\tau 3}^{}|^2 = 0 \; ,
\quad |\widetilde{U}_{e3}^{}|^2 = 1 \; ,
\nonumber \\
|\widetilde{U}_{\mu 1}^{}|^2  & =  |\widetilde{U}_{\tau 2}^{}|^2 =
\frac{1}{2} + \frac{\Delta_{21}^{} \left(
|U_{\tau 2}^{}|^2 - |U_{\mu 2}^{}|^2 \right) +
\Delta_{31}^{} \left(|U_{\tau 3}^{}|^2 - |U_{\mu 3}^{}|^2
\right)}{2 \sqrt{p^2 - 4q}} \; ,
\nonumber\\
|\widetilde{U}_{\mu 2}^{}|^2  & =  |\widetilde{U}_{\tau 1}^{}|^2 =
\frac{1}{2} - \frac{\Delta_{21}^{} \left(
|U_{\tau 2}^{}|^2 - |U_{\mu 2}^{}|^2 \right) +
\Delta_{31}^{} \left(|U_{\tau 3}^{}|^2 - |U_{\mu 3}^{}|^2
\right)}{2 \sqrt{p^2 - 4q}} \; .
\tag{28}
\end{align*}
To be more intuitive and instructive, let us take $\alpha \equiv \Delta^{}_{21}/
\Delta^{}_{31}$ and $|U_{e3}^{}|^2$ as two small expansion
parameters to simplify Eq. (28), because both of them are of
${\cal O}(10^{-2})$. Then we arrive at
\begin{align*}
|\widetilde{U}_{\mu 1}^{}|^2  & =  |\widetilde{U}_{\tau 2}^{}|^2 \simeq
1 - |{U}_{\mu 3}^{}|^2 \left(1+|{U}_{e 3}^{}|^2\right) +
\alpha\left(|{U}_{\mu 3}^{}|^2-|{U}_{\mu 3}^{}|^2|{U}_{e1}^{}|^2 -
|{U}_{\tau 1}^{}|^2\right) \; ,
\nonumber\\
|\widetilde{U}_{\mu 2}^{}|^2  & =  |\widetilde{U}_{\tau 1}^{}|^2 \simeq
|{U}_{\mu 3}^{}|^2 \left(1+|{U}_{e 3}^{}|^2\right) -
\alpha\left(|{U}_{\mu 3}^{}|^2-|{U}_{\mu 3}^{}|^2|{U}_{e1}^{}|^2 -
|{U}_{\tau 1}^{}|^2\right) \; .
\tag{29}
\end{align*}
It becomes clear that the asymptotic form of $\widetilde{U}$
for $A \to \infty$ contains only a single degree of freedom and
thus can be parametrized as
\begin{align*}
\left. \widetilde{U}\right|_{A\to \infty} =
\begin{pmatrix} 0 & 0 & 1 \cr \cos\theta & ~\sin\theta~ & 0 \cr
-\sin\theta & \cos\theta & 0 \end{pmatrix} \; ,
\tag{30}
\end{align*}
where
\begin{align*}
\tan\theta & =  \frac{\sqrt{p^2 - 4q} \ - \Delta_{21}^{} \left(
|U_{\tau 2}^{}|^2 - |U_{\mu 2}^{}|^2 \right) -
\Delta_{31}^{} \left(|U_{\tau 3}^{}|^2 - |U_{\mu 3}^{}|^2
\right)}{\sqrt{p^2 - 4q} \ + \Delta_{21}^{} \left(
|U_{\tau 2}^{}|^2 - |U_{\mu 2}^{}|^2 \right) +
\Delta_{31}^{} \left(|U_{\tau 3}^{}|^2 - |U_{\mu 3}^{}|^2
\right)}
\nonumber \\
& \simeq  \frac{|{U}_{\mu 3}^{}|^2 \left(1+|{U}_{e 3}^{}|^2 -
|{U}_{\mu 3}^{}|^2\right)}{\left(1 - |{U}_{\mu 3}^{}|^2 \right)^2}
+ \alpha \frac{|{U}_{\tau 1}^{}|^2-|{U}_{e 2}^{}|^2|{U}_{\mu 3}^{}|^2}
{\left(1 - |{U}_{\mu 3}^{}|^2 \right)^2} \; . \hspace{0.8cm}
\tag{31}
\end{align*}
It is easy to see that matter effects
{\it do} preserve the $\mu$-$\tau$ symmetry (i.e.,
$|\widetilde{U}^{}_{\mu i}| = |\widetilde{U}^{}_{\tau i}|$ will hold
as a consequence of $|U^{}_{\mu i}| = |U^{}_{\tau i}|$ for $i=1,2,3$)
even in very dense matter (i.e., $A \to \infty$).

At this point it is appropriate to clarify the confusing results obtained
before for $\widetilde{\theta}^{}_{12}$, $\widetilde{\theta}^{}_{23}$
and $\widetilde{\delta}$ in the $A \to \infty$ limit \cite{Wang:2019yfp,Xing:2018lob}.
Given $\widetilde{\theta}^{}_{13} \to \pi/2$ in this case, Eq. (25)
becomes
\begin{align*}
\left. \widetilde{U}\right|_{A \to \infty} & =  \left(\begin{matrix}
0 & 0 & e^{-{\rm i} \widetilde{\delta}}
\cr -\widetilde{s}^{}_{12} \widetilde{c}^{}_{23} - \widetilde{c}^{}_{12}
\widetilde{s}^{}_{23} e^{{\rm i} \widetilde{\delta}} &
\widetilde{c}^{}_{12} \widetilde{c}^{}_{23} -
\widetilde{s}^{}_{12} \widetilde{s}^{}_{23}
e^{{\rm i} \widetilde{\delta}} & 0 \cr \widetilde{s}^{}_{12}
\widetilde{s}^{}_{23} - \widetilde{c}^{}_{12} \widetilde{c}^{}_{23}
e^{{\rm i} \widetilde{\delta}} & ~ - \widetilde{c}^{}_{12} \widetilde{s}^{}_{23}
- \widetilde{s}^{}_{12} \widetilde{c}^{}_{23}
e^{{\rm i} \widetilde{\delta}} ~ & 0 \cr
\end{matrix} \right)
\nonumber \\
& =  \begin{pmatrix}
e^{-{\rm i} \widetilde \delta} & 0 & ~0 \cr
0 & e^{{\rm i} \widetilde \delta} & ~0 \cr 0 & 0 & ~1 \end{pmatrix}
\left(\begin{matrix}
0 & 0 &~ 1 \cr
X e^{{\rm i} \varphi}& Y e^{-{\rm i} \phi} &~ 0 \cr
-Y e^{{\rm i} \phi}& X e^{-{\rm i} \varphi} &~ 0
\end{matrix} \right)
\nonumber \\
& =  \begin{pmatrix}
e^{-{\rm i} \widetilde{\delta}} & 0 & 0 \cr
0 & e^{{\rm i}  (\widetilde\delta - \phi) } & 0 \cr
0 & 0 & e^{-{\rm i} \varphi} \end{pmatrix}
\begin{pmatrix}
0 & 0 & 1 \cr X & Y & 0 \cr -Y & X & 0 \end{pmatrix}
\begin{pmatrix}
e^{{\rm i} \left(\varphi + \phi\right)} & 0 & ~0 \cr
0 & 1 & ~0 \cr 0 & 0 & ~1 \end{pmatrix} \; , \hspace{0.8cm}
\tag{32}
\end{align*}
where $X \equiv |\widetilde{c}^{}_{12} \widetilde{s}^{}_{23} + \widetilde{s}^{}_{12}
\widetilde{c}^{}_{23} e^{-{\rm i} \widetilde{\delta}}|$,
$Y = |\widetilde{s}^{}_{12} \widetilde{s}^{}_{23} - \widetilde{c}^{}_{12}
\widetilde{c}^{}_{23} e^{-{\rm i} \widetilde{\delta}}|$,
$\varphi \equiv \pi + \arg(\widetilde{c}^{}_{12} \widetilde{s}^{}_{23} +
\widetilde{s}^{}_{12} \widetilde{c}^{}_{23} e^{-{\rm i} \widetilde{\delta}})$
and $\phi \equiv \pi+ \arg(\widetilde{s}^{}_{12} \widetilde{s}^{}_{23} -
\widetilde{c}^{}_{12} \widetilde{c}^{}_{23} e^{{\rm i} \widetilde{\delta}})$.
Then Eq. (32) is equivalent to Eq. (30) for the following
two reasons: first, the two diagonal phase matrices in Eq. (32) can be absorbed by
redefining the phases of the charged-lepton and neutrino fields
\footnote{Since neutrino oscillations are completely insensitive to the Majorana
phases of three massive neutrinos no matter whether matter effects are involved
or not, this rephasing treatment of the neutrino fields is definitely allowed
in this connection.};
second, $X^2 + Y^2 =1$ holds, and thus one may always take $X = \cos\theta$ and
$Y = \sin\theta$ with $\theta$ being in the first quadrant.

The above discussion implies that the individual values of $\widetilde{\theta}^{}_{12}$
and $\widetilde{\theta}^{}_{23}$ in the $A \to \infty$ limit do not make much sense,
and in particular the finite value of $\widetilde{\delta}$ in this case is misleading.
The latter observation is also supported by the fact $\widetilde{\cal J} \to
0$ for $A \to \infty$, as guaranteed by the Naumov relation in Eq. (24). A question
turns out to be why $\widetilde{\cal J}$ and $|\widetilde{U}^{}_{\alpha i}|^2$ have
the well-defined asymptotic behaviors in very dense matter, but the parameters
$\widetilde{\theta}^{}_{12}$, $\widetilde{\theta}^{}_{13}$, $\widetilde{\theta}^{}_{23}$
and $\widetilde{\delta}$ may not have. The answer to this question is very simple:
a specific parametrization of $\widetilde{U}$ is always basis-dependent and hence
its parameters are not guaranteed to be fully physical in the extreme case in
which a redefinition of the basis becomes available to remove the possible
parameter redundancy. In contrast, $\widetilde{\cal J}$ and
$|\widetilde{U}^{}_{\alpha i}|^2$ do not suffer from this kind of subtlety
because they are rephasing-invariant or basis-independent.

\subsection{Case B: (IMO, $\nu$)}

Now we turn to the case of a neutrino beam ($\nu$)
propagating in matter with an inverted mass ordering (IMO).
In the $A \to \infty$ limit, Eq. (4) is reduced to
\begin{align*}
& \widetilde{\Delta}^{}_{21} = \Delta^{}_{21} + \Delta^{}_{31} + A -
\frac{1}{2}\left(3 p + \sqrt{p^2 - 4q}\right) \;,
\nonumber \\
& \widetilde{\Delta}^{}_{31} = -\sqrt{p^2 - 4 q}  \;, \hspace{1cm}
\nonumber \\
& B = \frac{1}{2} \left(p + \sqrt{p^2 - 4 q}\right) \;,
\tag{33}
\end{align*}
where $p$ and $q$ have already been defined below Eq. (27). In this case
we find that $\widetilde \Delta_{31}^{} \simeq \Delta_{31}^{}
\left(1-|U_{e3}^{}|^2\right) - \Delta^{}_{21} |U_{e1}^{}|^2$
is finite and $\widetilde \Delta_{21}^{} \simeq -
\widetilde \Delta_{32}^{} \simeq A$ approaches infinity when
$A \to \infty$ is taken for very dense matter.

With the help of Eq. (33) and Eqs. (13)---(15), we calculate
the nine elements of $\widetilde{U}$ in the $A \to \infty$
limit and get
\begin{align*}
|\widetilde{U}_{e1}^{}|^2  & =  |\widetilde{U}_{e3}^{}|^2 =
|\widetilde{U}_{\mu 2}^{}|^2 = |\widetilde{U}_{\tau 2}^{}|^2 = 0 \; ,
\quad |\widetilde{U}_{e2}^{}|^2 = 1 \; ,
\nonumber \\
|\widetilde{U}_{\mu 1}^{}|^2  & =  |\widetilde{U}_{\tau 3}^{}|^2 =
\frac{1}{2} - \frac{\Delta_{21}^{} \left(
	|U_{\tau 2}^{}|^2 - |U_{\mu 2}^{}|^2 \right) +
	\Delta_{31}^{} \left(|U_{\tau 3}^{}|^2 - |U_{\mu 3}^{}|^2
	\right)}{2 \sqrt{p^2 - 4q}} \; ,
\nonumber\\
|\widetilde{U}_{\mu 3}^{}|^2  & =  |\widetilde{U}_{\tau 1}^{}|^2 =
\frac{1}{2} + \frac{\Delta_{21}^{} \left(
	|U_{\tau 2}^{}|^2 - |U_{\mu 2}^{}|^2 \right) +
	\Delta_{31}^{} \left(|U_{\tau 3}^{}|^2 - |U_{\mu 3}^{}|^2
	\right)}{2 \sqrt{p^2 - 4q}} \; . \hspace{0.8cm}
\tag{34}
\end{align*}
Just as Eq. (29), the expressions of $|\widetilde{U}_{\mu 1}^{}|^2$,
$|\widetilde{U}_{\mu 3}^{}|^2$, $|\widetilde{U}_{\tau 1}^{}|^2$
and $|\widetilde{U}_{\tau 3}^{}|^2$ in Eq. (34) can be expanded in terms
of $\alpha$ and $|U_{e3}^{}|^2$. As a result,
\begin{align*}
|\widetilde{U}_{\mu 1}^{}|^2  & =  |\widetilde{U}_{\tau 3}^{}|^2 \simeq
1 - |{U}_{\mu 3}^{}|^2 \left(1+|{U}_{e 3}^{}|^2\right) +
\alpha\left(|{U}_{\mu 3}^{}|^2-|{U}_{\mu 3}^{}|^2|{U}_{e1}^{}|^2 -
|{U}_{\tau 1}^{}|^2\right) \;,\nonumber\\
|\widetilde{U}_{\mu 3}^{}|^2  & =  |\widetilde{U}_{\tau 1}^{}|^2
\simeq |{U}_{\mu 3}^{}|^2 \left(1+|{U}_{e 3}^{}|^2\right) -
\alpha\left(|{U}_{\mu 3}^{}|^2-|{U}_{\mu 3}^{}|^2|{U}_{e1}^{}|^2 -
|{U}_{\tau 1}^{}|^2\right) \; .
\tag{35}
\end{align*}
In this case the asymptotic form of $\widetilde{U}$ also contains
only a single degree of freedom and thus can be rewritten as
\begin{align*}
\left. \widetilde{U}\right|_{A\to \infty} =
\begin{pmatrix} 0 & 1 & 0 \cr \cos\theta & ~0~ & \sin\theta \cr
-\sin\theta & 0 & \cos\theta \end{pmatrix} \; ,
\tag{36}
\end{align*}
where
\begin{align*}
\tan\theta & =  \frac{\sqrt{p^2 - 4q} \ + \Delta_{21}^{} \left(
	|U_{\tau 2}^{}|^2 - |U_{\mu 2}^{}|^2 \right) +
	\Delta_{31}^{} \left(|U_{\tau 3}^{}|^2 - |U_{\mu 3}^{}|^2
	\right)}{\sqrt{p^2 - 4q} \ - \Delta_{21}^{} \left(
	|U_{\tau 2}^{}|^2 - |U_{\mu 2}^{}|^2 \right) -
	\Delta_{31}^{} \left(|U_{\tau 3}^{}|^2 - |U_{\mu 3}^{}|^2
	\right)}
\nonumber \\
& \simeq  \frac{|{U}_{\mu 3}^{}|^2 \left(1+|{U}_{e 3}^{}|^2 -
|{U}_{\mu 3}^{}|^2\right)}{\left(1 -
|{U}_{\mu 3}^{}|^2 \right)^2} + \alpha \frac{|{U}_{\tau 1}^{}|^2 -
|{U}_{e 2}^{}|^2|{U}_{\mu 3}^{}|^2}
{\left(1 - |{U}_{\mu 3}^{}|^2 \right)^2} \; . \hspace{0.8cm}
\tag{37}
\end{align*}

\subsection{Case C: (NMO, $\overline\nu$)}

When it comes to the case of an antineutrino beam
($\overline\nu$) propagating in matter with a normal
mass ordering (NMO), we simplify Eq. (3) in the
$A \to \infty$ limit as follows:
\begin{align*}
& \widetilde{\Delta}^{}_{21} = -\Delta^{}_{21} - \Delta^{}_{31} + A +
\frac{1}{2}\left(3 p - \sqrt{p^2 - 4q}\right) \; ,
\nonumber \\
& \widetilde{\Delta}^{}_{31} =-\Delta^{}_{21} - \Delta^{}_{31} + A +
\frac{1}{2}\left(3 p + \sqrt{p^2 - 4q}\right)  \; , \hspace{1cm}
\nonumber \\
& B = |U_{e2}^{}|^2 \Delta_{21}^{} + |U_{e3}^{}|^2 \Delta_{31}^{} -A \; ,
\tag{38}
\end{align*}
with $p$ and $q$ having been defined below Eq. (27). In this case
we find that $\widetilde \Delta_{32}^{} =
\sqrt{p^2 - 4q}\simeq \Delta_{31}^{}
\left(1-|U_{e3}^{}|^2\right) - \Delta^{}_{21} |U_{e1}^{}|^2$ is finite
and $\widetilde \Delta_{21}^{} \simeq \widetilde \Delta_{31}^{} \simeq A$
approaches infinity.

Given Eq. (38) and Eqs. (13)---(15), the elements of $\widetilde{U}$ in the
$A \to \infty$ limit read as
\begin{align*}
|\widetilde{U}_{e2}^{}|^2  & =  |\widetilde{U}_{e3}^{}|^2 =
|\widetilde{U}_{\mu 1}^{}|^2 = |\widetilde{U}_{\tau 1}^{}|^2 = 0 \; ,
\quad |\widetilde{U}_{e1}^{}|^2 = 1 \; ,
\nonumber \\
|\widetilde{U}_{\mu 2}^{}|^2  & =  |\widetilde{U}_{\tau 3}^{}|^2 =
\frac{1}{2} + \frac{\Delta_{21}^{} \left(
	|U_{\tau 2}^{}|^2 - |U_{\mu 2}^{}|^2 \right) +
	\Delta_{31}^{} \left(|U_{\tau 3}^{}|^2 - |U_{\mu 3}^{}|^2
	\right)}{2 \sqrt{p^2 - 4q}} \; ,
\nonumber\\
|\widetilde{U}_{\mu 3}^{}|^2  & =  |\widetilde{U}_{\tau 2}^{}|^2 =
\frac{1}{2} - \frac{\Delta_{21}^{} \left(
	|U_{\tau 2}^{}|^2 - |U_{\mu 2}^{}|^2 \right) +
	\Delta_{31}^{} \left(|U_{\tau 3}^{}|^2 - |U_{\mu 3}^{}|^2
	\right)}{2 \sqrt{p^2 - 4q}} \; . \hspace{0.8cm}
\tag{39}
\end{align*}
Expanding $|\widetilde{U}_{\mu 2}^{}|^2$,
$|\widetilde{U}_{\mu 3}^{}|^2$, $|\widetilde{U}_{\tau 2}^{}|^2$
and $|\widetilde{U}_{\tau 3}^{}|^2$ in terms of
$\alpha$ and
$|U_{e3}^{}|^2$, we obtain
\begin{align*}
|\widetilde{U}_{\mu 2}^{}|^2  & =  |\widetilde{U}_{\tau 3}^{}|^2 \simeq
1 - |{U}_{\mu 3}^{}|^2 \left(1+|{U}_{e 3}^{}|^2\right) +
\alpha\left(|{U}_{\mu 3}^{}|^2-|{U}_{\mu 3}^{}|^2|{U}_{e1}^{}|^2 -
|{U}_{\tau 1}^{}|^2\right) \;,\nonumber\\
|\widetilde{U}_{\mu 3}^{}|^2  & =  |\widetilde{U}_{\tau 2}^{}|^2 
\simeq |{U}_{\mu 3}^{}|^2 \left(1+|{U}_{e 3}^{}|^2\right) -
\alpha\left(|{U}_{\mu 3}^{}|^2-|{U}_{\mu 3}^{}|^2|{U}_{e1}^{}|^2 -
|{U}_{\tau 1}^{}|^2\right) \; .
\tag{40}
\end{align*}
In this case the asymptotic form of $\widetilde{U}$ can be
parameterized with only a single degree of freedom as follows:
\begin{align*}
\left. \widetilde{U}\right|_{A\to \infty} =
\begin{pmatrix} 1 & 0 & 0 \cr 0 & \cos\theta & \sin\theta \cr
0 & -\sin\theta & \cos\theta \end{pmatrix} \; ,
\tag{41}
\end{align*}
where
\begin{align*}
\tan\theta & =  \frac{\sqrt{p^2 - 4q} \ - \Delta_{21}^{} \left(
|U_{\tau 2}^{}|^2 - |U_{\mu 2}^{}|^2 \right) -
\Delta_{31}^{} \left(|U_{\tau 3}^{}|^2 - |U_{\mu 3}^{}|^2
\right)}{\sqrt{p^2 - 4q} \ + \Delta_{21}^{} \left(
|U_{\tau 2}^{}|^2 - |U_{\mu 2}^{}|^2 \right) +
\Delta_{31}^{} \left(|U_{\tau 3}^{}|^2 - |U_{\mu 3}^{}|^2
\right)} \nonumber \\
& \simeq \frac{|{U}_{\mu 3}^{}|^2 \left(1+|{U}_{e 3}^{}|^2 -
|{U}_{\mu 3}^{}|^2\right)}{\left(1 - |{U}_{\mu 3}^{}|^2 \right)^2}
+ \alpha \frac{|{U}_{\tau 1}^{}|^2-|{U}_{e 2}^{}|^2|{U}_{\mu 3}^{}|^2}
{\left(1 - |{U}_{\mu 3}^{}|^2 \right)^2} \; .
\tag{42}
\end{align*}

\subsection{Case D: (IMO, $\overline\nu$)}

Similarly, in the case of an anti-neutrino beam ($\overline \nu$)
propagating in matter with an inverted mass ordering (IMO),
we simplify Eq. (4) in the $A \to \infty$ limit and arrive at
\begin{align*}
& \widetilde{\Delta}^{}_{21} = \sqrt{p^2 - 4q} \;\; ,
\nonumber \\
& \widetilde{\Delta}^{}_{31} =  \Delta^{}_{21} + \Delta^{}_{31} - A -
\frac{1}{2}\left(3 p - \sqrt{p^2 - 4q}\right) \; , \hspace{1cm}
\nonumber \\
& B = \frac{1}{2} \left(p - \sqrt{p^2 - 4 q}\right) \; ,
\tag{43}
\end{align*}
where $p$ and $q$ have been given below Eq. (27). We are therefore left
with finite $\widetilde \Delta_{21}^{} \simeq -\Delta_{31}^{}
\left(1-|U_{e3}^{}|^2\right) + \Delta^{}_{21} |U_{e1}^{}|^2$
and infinite $\widetilde \Delta_{31}^{} \simeq
\widetilde \Delta_{32}^{} \simeq -A$. The expressions of
$|\widetilde{U}_{\alpha i}^{}|^2$ in the $A \to \infty$
limit turn out to be
\begin{align*}
|\widetilde{U}_{e1}^{}|^2  & =  |\widetilde{U}_{e2}^{}|^2 =
|\widetilde{U}_{\mu 3}^{}|^2 = |\widetilde{U}_{\tau 3}^{}|^2 = 0 \; ,
\quad |\widetilde{U}_{e3}^{}|^2 = 1 \; ,
\nonumber \\
|\widetilde{U}_{\mu 1}^{}|^2  & =  |\widetilde{U}_{\tau 2}^{}|^2 =
\frac{1}{2} + \frac{\Delta_{21}^{} \left(
|U_{\tau 2}^{}|^2 - |U_{\mu 2}^{}|^2 \right) +
\Delta_{31}^{} \left(|U_{\tau 3}^{}|^2 - |U_{\mu 3}^{}|^2
\right)}{2 \sqrt{p^2 - 4q}} \; ,
\nonumber\\
|\widetilde{U}_{\mu 2}^{}|^2  & =  |\widetilde{U}_{\tau 1}^{}|^2 =
\frac{1}{2} - \frac{\Delta_{21}^{} \left(
|U_{\tau 2}^{}|^2 - |U_{\mu 2}^{}|^2 \right) +	
\Delta_{31}^{} \left(|U_{\tau 3}^{}|^2 - |U_{\mu 3}^{}|^2
\right)}{2 \sqrt{p^2 - 4q}} \; . \hspace{0.8cm}
\tag{44}
\end{align*}
In a good approximation, we find
\begin{align*}
|\widetilde{U}_{\mu 1}^{}|^2  & =  |\widetilde{U}_{\tau 2}^{}|^2
\simeq |{U}_{\mu 3}^{}|^2 \left(1+|{U}_{e 3}^{}|^2\right) -
\alpha\left(|{U}_{\mu 3}^{}|^2-|{U}_{\mu 3}^{}|^2|{U}_{e1}^{}|^2 -
|{U}_{\tau 1}^{}|^2\right)  \;,\nonumber\\
|\widetilde{U}_{\mu 2}^{}|^2  & =  |\widetilde{U}_{\tau 1}^{}|^2
\simeq 1 - |{U}_{\mu 3}^{}|^2 \left(1+|{U}_{e 3}^{}|^2\right) +
\alpha\left(|{U}_{\mu 3}^{}|^2-|{U}_{\mu 3}^{}|^2|{U}_{e1}^{}|^2 -
|{U}_{\tau 1}^{}|^2\right)\; .
\tag{45}
\end{align*}
In this case the asymptotic form of $\widetilde{U}$ contains only a single degree
of freedom and can be parametrized as
\begin{align*}
\left. \widetilde{U}\right|_{A\to \infty} =
\begin{pmatrix} 0 & 0 & ~1 \cr \cos\theta & \sin\theta & ~0 \cr
-\sin\theta & \cos\theta & ~0 \end{pmatrix} \; ,
\tag{46}
\end{align*}
where
\begin{align*}
\tan\theta & =  \frac{\sqrt{p^2 - 4q} \ - \Delta_{21}^{} \left(
|U_{\tau 2}^{}|^2 - |U_{\mu 2}^{}|^2 \right) -
\Delta_{31}^{} \left(|U_{\tau 3}^{}|^2 - |U_{\mu 3}^{}|^2
\right)}{\sqrt{p^2 - 4q} \ + \Delta_{21}^{} \left(
|U_{\tau 2}^{}|^2 - |U_{\mu 2}^{}|^2 \right) +
\Delta_{31}^{} \left(|U_{\tau 3}^{}|^2 - |U_{\mu 3}^{}|^2
\right)} \nonumber \\
& \simeq  \frac{|{U}_{\tau 3}^{}|^2}{|{U}_{\mu 3}^{}|^2} -
\alpha  \frac{|{U}_{\tau 1}^{}|^2
-|{U}_{e 2}^{}|^2|{U}_{\mu 3}^{}|^2 }{ |{U}_{\mu 3}^{}|^4 } \; .
\hspace{0.8cm}
\tag{47}
\end{align*}
Comparing between cases A and B (or between C and D), we immediately
find the interesting relations
\begin{align*}
&\left.|\widetilde{U}_{\mu 1}^{}|^2\right|_{ A \to
\infty}^{({{\rm NMO},~\nu})} =
\left.|\widetilde{U}_{\mu 2}^{}|^2\right|_{ A \to
\infty}^{({{\rm NMO},~\overline\nu})} \; ,
\nonumber \\
&\left.|\widetilde{U}_{\mu
1}^{}|^2\right|_{ A \to \infty}^{({{\rm IMO},~\nu})}
=\left.|\widetilde{U}_{\mu 2}^{}|^2\right|_{ A \to
\infty}^{({{\rm IMO},~ \overline\nu})} \; ,
\tag{48}
\end{align*}
which are equivalent to $\left.\tan\theta \right|_{ A \to
\infty}^{({{\rm NMO},~\nu})}= \left. \tan\theta \right|_{ A \to
\infty}^{({{\rm NMO},~\overline\nu})}$ and
$\left.\tan\theta \right|_{ A \to \infty}^{({{\rm
IMO},~\nu})}=\left.\cot\theta \right|_{ A \to
\infty}^{({{\rm IMO},~ \overline\nu})}$.

Taking account of the best-fit values of the six neutrino oscillation
parameters (i.e., $\Delta_{21}^{}=7.39 \times 10^{-5} ~{\rm eV}^2$,
$\Delta_{31}^{}=2.525 \times 10^{-3} ~{\rm eV}^2$,
$\theta_{12}^{} = 33.82^{\circ}$, $\theta_{13}^{} =
8.61^{\circ}$, $\theta_{23}^{} = 49.7^{\circ}$ and
$\delta=217^{\circ}$ for the NMO case; or $\Delta_{21}^{}=7.39
\times 10^{-5} ~{\rm eV}^2$, $\Delta_{31}^{}=-2.438 \times 10^{-3} ~{\rm eV}^2$,
$\theta_{12}^{} = 33.82^{\circ}$, $\theta_{13}^{} =
8.65^{\circ}$, $\theta_{23}^{} = 49.7^{\circ}$ and
$\delta=280^{\circ}$ for the IMO case) \cite{Esteban:2018azc,Capozzi:2018ubv},
we have
\begin{align*}
\begin{pmatrix}
|U_{e1}^{}|^2 & |U_{e2}^{}|^2 & |U_{e3}^{}|^2 \cr
|U_{\mu 1}^{}|^2 & |U_{\mu 2}^{}|^2 & |U_{\mu 3}^{}|^2 \cr
|U_{\tau 1}^{}|^2 & |U_{\tau 2}^{}|^2 & |U_{\tau 3}^{}|^2
\end{pmatrix}\simeq
\begin{pmatrix}
0.675 & 0.303 & 0.022 \cr
0.084 & 0.347 & 0.569 \cr
0.241 & 0.350 & 0.409
\end{pmatrix}
\tag{49}
\end{align*}
for the NMO case; or
\begin{align*}
\begin{pmatrix}
|U_{e1}^{}|^2 & |U_{e2}^{}|^2 & |U_{e3}^{}|^2 \cr
|U_{\mu 1}^{}|^2 & |U_{\mu 2}^{}|^2 & |U_{\mu 3}^{}|^2 \cr
|U_{\tau 1}^{}|^2 & |U_{\tau 2}^{}|^2 & |U_{\tau 3}^{}|^2
\end{pmatrix}\simeq
\begin{pmatrix}
0.674 & 0.303 & 0.023 \cr
0.151 & 0.281 & 0.568 \cr
0.175 & 0.416 & 0.409
\end{pmatrix}
\tag{50}
\end{align*}
for the IMO case. Starting from the best-fit values of
$|U_{\alpha i}|^2$ at $A = 0$ as given in Eq. (49) and Eq. (50),
each of the nine effective quantities
$|\widetilde U_{\alpha i}|^2$ evolves with the matter parameter $A$ in a way
shown in Figs. 1 and 2, where case A (NMO, $\nu$),
case B (IMO, ${\nu}$), case C (NMO, $\overline\nu$) and
case D (IMO, $\overline{\nu}$) have all been taken into
account. One can see that $|\widetilde U_{\alpha i}|^2 \simeq |U_{\alpha i}|^2$
is a good approximation when $A$ is small enough (i.e.,
$A \lesssim 10^{-6} ~ {\rm eV}^2$).
If the matter parameter $A$ lies in the range
$10^{-6} ~ {\rm eV}^2 \lesssim A \lesssim 10^{-2} ~ {\rm eV}^2$,
matter effects turn out to be significant and can make important corrections to the
genuine lepton flavor mixing matrix $U$ in vacuum. When taking the value of
$A$ larger than $10^{-2} ~{\rm eV}^2$, we find that the effective PMNS matrix
$\widetilde{U}$ asymptotically approaches a constant matrix in the $A \to \infty$
limit. To be more explicit, we obtain
\begin{figure}[h]
\centering{
\includegraphics[]{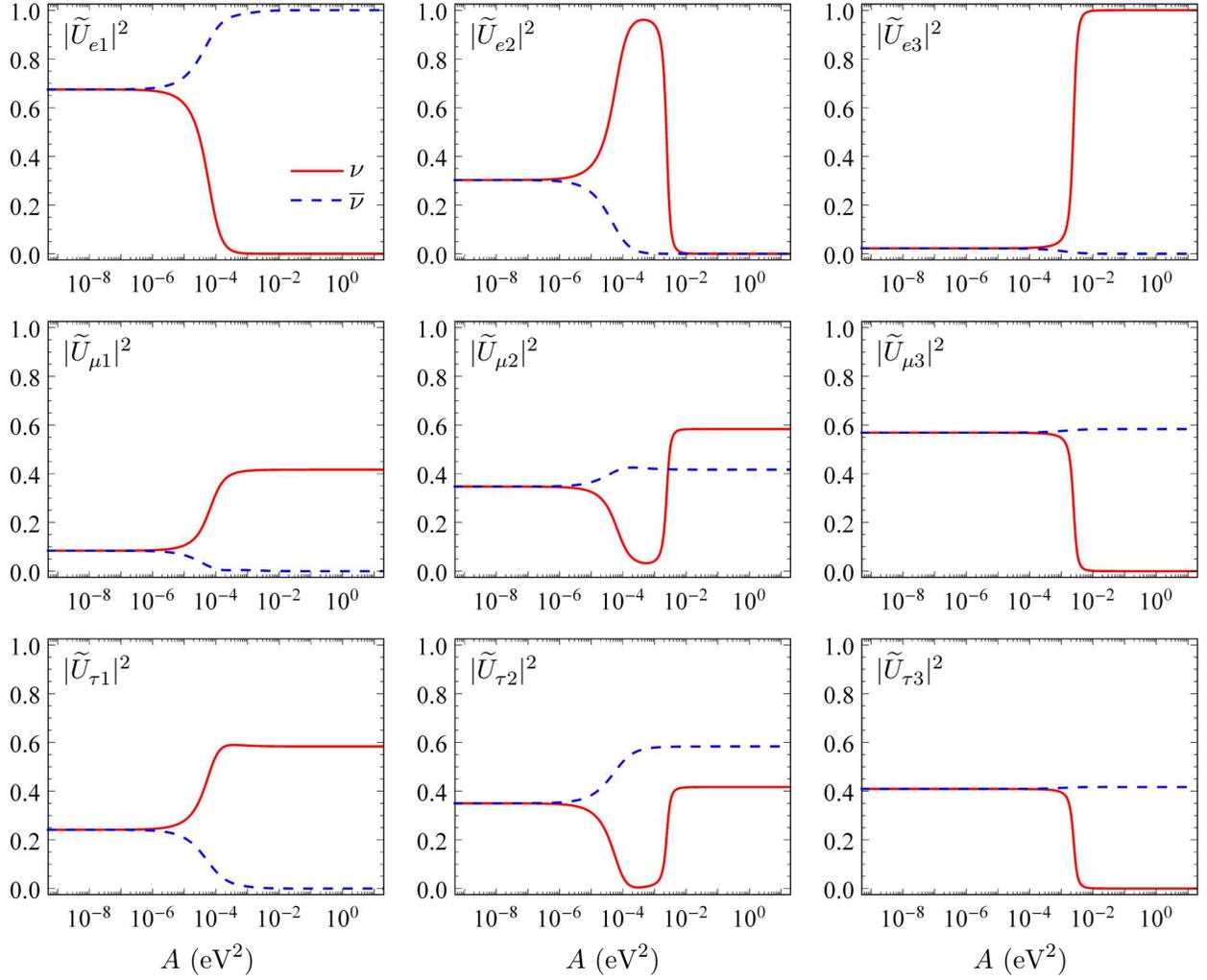}}
\caption{The evolution of $|\widetilde U_{\alpha i}^{}|^2$ (for $\alpha=e,\mu,\tau$
and $i=1,2,3$) with the matter effect parameter $A$ in the
normal neutrino mass ordering case, where the best-fit
values of six neutrino oscillation parameters have been input
\cite{Esteban:2018azc}.}
\end{figure}
\begin{align*}
\left. \left(|\widetilde{U}_{\alpha i}^{}|^2\right)
\right|_{A\to \infty}^{({{\rm NMO},~\nu})} =
\begin{pmatrix} 0 & 0 & ~1 \cr 0.417 & 0.583 & ~0 \cr
0.583 & 0.417 & ~0 \end{pmatrix} \; ,
\tag{51a}
\end{align*}
\vspace{-0.6 cm}
\begin{align*}
\left. \left(|\widetilde{U}_{\alpha i}^{}|^2\right)
\right|_{A\to \infty}^{({{\rm NMO},
~\overline\nu})} = \begin{pmatrix} 1~ & 0 & 0 \cr 0~ & 0.417 &
0.583 \cr 0~ & 0.583 & 0.417\end{pmatrix} \; ,
\tag{51b}
\end{align*}
\vspace{-0.6 cm}
\begin{align*}
\left.\left(|\widetilde{U}_{\alpha i}^{}|^2\right)
\right|_{A\to \infty}^{({{\rm IMO},~\nu})} =
\begin{pmatrix} 0 & ~1~ & 0 \cr 0.418 & ~0~ & 0.582  \cr
0.582 & ~0~ &  0.418 \end{pmatrix} \; ,
\tag{51c}
\end{align*}
\vspace{-0.6 cm}
\begin{align*}
\left. \left(|\widetilde{U}_{\alpha i}^{}|^2\right)
\right|_{A\to \infty}^{({{\rm IMO},
~\overline\nu})} = \begin{pmatrix} 0 & 0 & ~1 \cr
0.582 & 0.418 & ~0 \cr 0.418 & 0.582 & ~0 \end{pmatrix} \; .
\tag{51d}
\end{align*}
If the $3 \sigma$ ranges of six neutrino oscillation parameters
\cite{Esteban:2018azc} are taken into account, we get
\begin{align*}
&\left.|\widetilde{U}_{\mu 1}^{}|^2\right|_{A\to
\infty}^{({{\rm NMO},~\nu})}=\left.|\widetilde{U}_{\mu
2}^{}|^2\right|_{A\to \infty}^{({{\rm NMO},~\overline\nu})}
= 0.373 \to 0.574 \; ,
\nonumber \\
&\left.|\widetilde{U}_{\mu
1}^{}|^2\right|_{A\to \infty}^{({{\rm IMO},~\nu})}
=\left.|\widetilde{U}_{\mu 2}^{}|^2\right|_{A\to
\infty}^{({{\rm IMO},~\overline\nu})} = 0.375 \to 0.569 \; .
\tag{52}
\end{align*}
\begin{figure}[h]
\centering{
\includegraphics[]{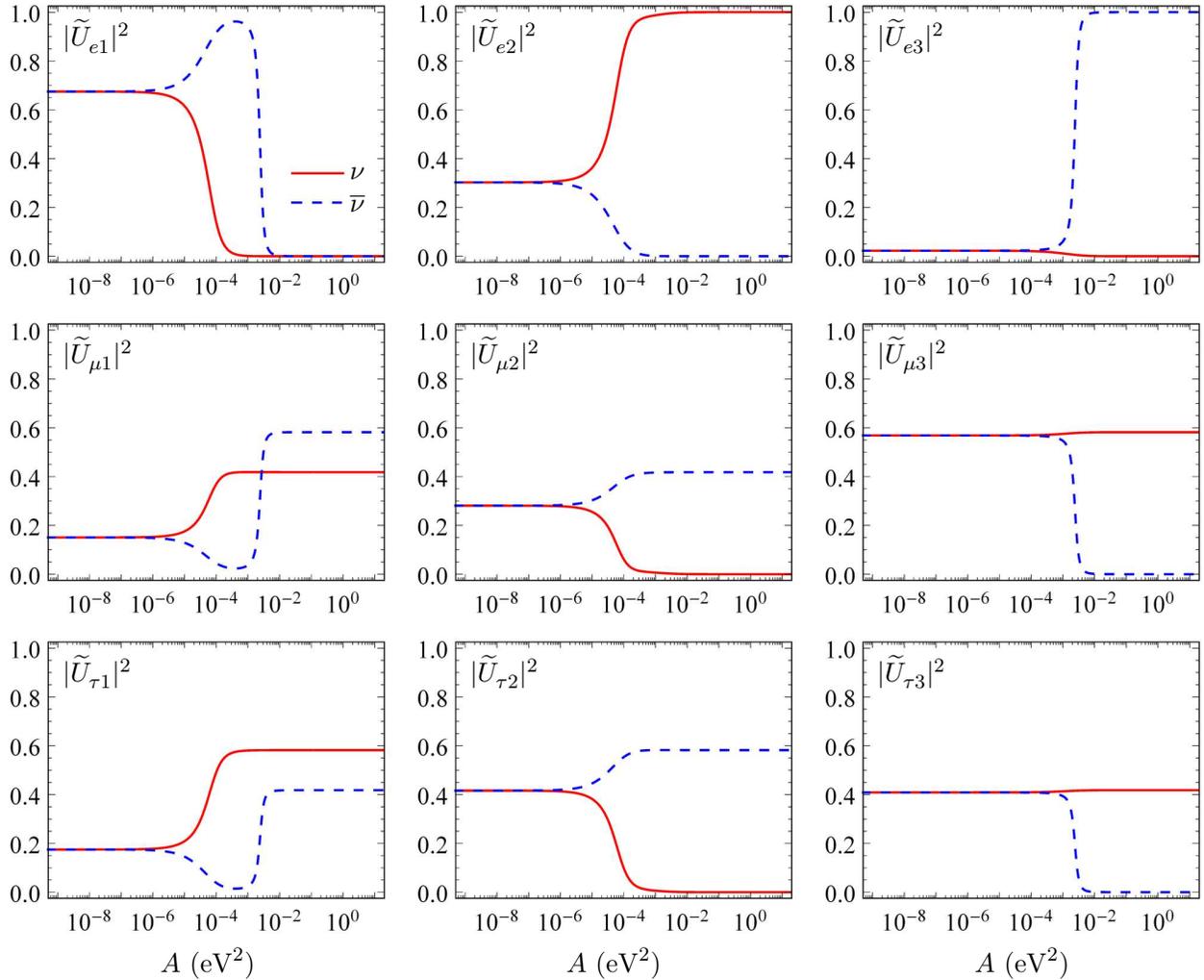}}
\caption{The evolution of $|\widetilde U_{\alpha i}^{}|^2$ (for $\alpha=e,\mu,\tau$
and $i=1,2,3$) with the matter effect parameter $A$ in the
inverted neutrino mass ordering case, where the best-fit
values of six neutrino oscillation parameters have been input
\cite{Esteban:2018azc}.}
\end{figure}

\section{Summary}

How to transparently describe matter effects on the
behaviors of neutrino oscillations has been
an interesting and important topic in neutrino phenomenology, and among a number of
useful approaches making use of the effective PMNS matrix $\widetilde{U}$ and the
effective neutrino mass-squared differences $\widetilde{\Delta}^{}_{ji}$ has proved
convenient to discuss neutrino mixing in a medium and formulate the matter-corrected
probabilities of flavor oscillations. Then it makes sense to explore possible
asymptotic behaviors of these effective quantities in the regime where the matter
density is sufficiently large, a case which is mathematically equivalent to assuming
the matter parameter $A = 2\sqrt{2} ~ G^{}_{\rm F} N^{}_e E$ to approach infinity.

In this paper we have established some direct and concise relations between
$(\widetilde{U}^{}_{\alpha i}, \widetilde{\Delta}^{}_{ji})$ in matter and their
fundamental counterparts $(U^{}_{\alpha i}, \Delta^{}_{ji})$ in vacuum (for
$\alpha =e, \mu, \tau$ and $i,j =1,2,3$) with the help of two sets of sum rules
for them. These sum rules allow us to derive new and exact formulas for both
nine $|\widetilde{U}^{}_{\alpha i}|^2$ and nine $\widetilde{U}^{}_{\alpha i}
\widetilde{U}^*_{\beta i}$ in a parametrization-independent way, by which we
have analytically unraveled the asymptotic behaviors of $|\widetilde{U}^{}_{\alpha i}|^2$
and $\widetilde{\Delta}^{}_{ji}$ in very dense matter for the first time.
We have also clarified the confusion associated with the parameter redundancy of $\widetilde{\theta}^{}_{12}$, $\widetilde{\theta}^{}_{13}$, $\widetilde{\theta}^{}_{23}$
and $\widetilde{\delta}$ in the standard parametrization of $\widetilde{U}$ in
the $A \to \infty$ limit. We conclude that $\widetilde{U}$ contains only a single
degree of freedom in this extreme case, with no CP violation in neutrino
oscillations.

Finally it is worth mentioning that our approach can easily be extended to the
(3+1) neutrino mixing scheme in matter, in which three active neutrinos are mixed
with one light sterile neutrino species denoted as $\nu^{}_s$ (flavor) or
$\nu^{}_4$ (mass). The corresponding sum rules for
$\widetilde{U}$ and $\widetilde{\Delta}^{}_{ji}$ will help
derive the exact formulas for sixteen
$|\widetilde{U}^{}_{\alpha i}|^2$ and sixteen $\widetilde{U}^{}_{\alpha i}
\widetilde{U}^*_{\beta i}$ (for $\alpha,\beta = e, \mu, \tau, s$ and
$i =1,2,3,4$), as previously done \cite{Zhang:2006yq}. In this case the matter
potential term proportional to $V^{}_{\rm nc}$ must be taken into account.

\vspace{0.5cm}

We would like to thank Shu Luo for her partial involvement and useful
discussions at the early stage of this work. We are also grateful to
Yu-Feng Li, Xin Wang, Di Zhang, Shun Zhou and Ye-Ling Zhou for helpful
discussions on matter effects. This work was supported in part by
the National Natural Science Foundation of China under grant No. 11775231
and grant No. 11835013.

\vspace{0.5cm}

\end{document}